\documentclass[manuscript]{aastex}
\usepackage{rotating}
\def\deg{\ifmmode^\circ\else$^\circ$\fi}

\def\arcs{\ifmmode {''}\else $''$\fi}
\def\arcm{\ifmmode {'}\else $'$\fi}
\def\parcs{\sa=.07em \sb=.03em
     \ifmmode $\rlap{.}$^{\scriptscriptstyle\prime\kern -\sb\prime}$\kern -\sa$
     \else \rlap{.}$^{\scriptscriptstyle\prime\kern -\sb\prime}$\kern -\sa\fi}
\def\parcm{\sa=.08em \sb=.03em
     \ifmmode $\rlap{.}\kern\sa$^{\scriptscriptstyle\prime}$\kern-\sb$
     \else \rlap{.}\kern\sa$^{\scriptscriptstyle\prime}$\kern-\sb\fi}

\def\spose#1{\hbox to 0pt{#1\hss}}
\def\simlt{\mathrel{\spose{\lower 3pt\hbox{$\mathchar"218$}}
     \raise 2.0pt\hbox{$\mathchar"13C$}}}
\def\simgt{\mathrel{\spose{\lower 3pt\hbox{$\mathchar"218$}}
     \raise 2.0pt\hbox{$\mathchar"13E$}}}
\def\lsim{\rlap{$<$}{\lower 1.0ex\hbox{$\sim$}}}
\def\gsim{\rlap{$>$}{\lower 1.0ex\hbox{$\sim$}}}
\usepackage{epsf}
\usepackage{psfig}
\usepackage [english]{babel}
\usepackage [autostyle, english = american]{csquotes}
\MakeOuterQuote{"}
\usepackage{gensymb}

\begin{document}

\title{Accretion-Inhibited Star formation in the Warm Molecular Disk of the Green-valley Elliptical Galaxy NGC 3226? } 

\author{P. N. Appleton\altaffilmark{1}, C. Mundell\altaffilmark{2}, T. Bitsakis\altaffilmark{1,3}, M. Lacy\altaffilmark{4}, K. Alatalo\altaffilmark{1}, L. Armus\altaffilmark{5}, V. Charmandaris\altaffilmark{6,7,8}, P-A. Duc\altaffilmark{9}, U. Lisenfeld\altaffilmark{10}, P. Ogle\altaffilmark{11}}

\altaffiltext{1}{NASA {\it Herschel} Science Center, Infrared Processing and Analysis Center, Caltech, 770S Wilson Av., Pasadena, CA 91125. apple@ipac.caltech.edu}
\altaffiltext{2}{Astrophysics Research Institute,  John Moores University, Liverpool Science Park, 146 Brownlow Hill, Liverpool L3 5RF, UK}
\altaffiltext{3}{Instituto de Astronomia, National Autonomous University of Mexico, P. O. 70-264, 04510 D. F., Mexico}
\altaffiltext{4}{NRAO, Charlottesville}
\altaffiltext{5}{{\it Spitzer} NASA Herschel Science Center, 1200 E. California Blvd., Caltech, Pasadena, CA 91125}
\altaffiltext{6}{Department of Physics, University of Crete, GR-71003, Heraklion, Greece}
\altaffiltext{7}{Institute for Astronomy, Astrophysics, Space Applications \& Remote Sensing, National Observatory of Athens, GR-15236, Penteli, Greece}
\altaffiltext{8}{Chercheur Associ\'e, Observatoire de Paris, F-75014, Paris, France}
\altaffiltext{9}{Service d'Astrophysique, Laboratoire AIM, CEA-Saclay, Orme des Merisiers, Bat 709, 91191 Gif sur Yvette , France}
\altaffiltext{10}{Dept. Fisica Teorica y del Cosmos, University of Granada, Edifica Mecenas,Granada, Spain}
\altaffiltext{11}{NASA Extragalactic Database, IPAC, Caltech, 1200 E. California Blvd, Caltech, Pasadena, CA 91125}

\begin{abstract}

We present archival {\it Spitzer} photometry and spectroscopy, and
{\it Herschel} photometry, of the peculiar "Green Valley" elliptical galaxy NGC~3226. The galaxy, which contains a low-luminosity AGN,  forms a pair with NGC~3227, and is shown to lie in a complex web of stellar and HI filaments. Imaging at 8 and 16$\mu$m reveals a curved plume structure 3 kpc in extent, embedded within the core of the galaxy, and coincident with the termination of a 30 kpc-long HI tail. In-situ star formation associated with the IR plume is identified from narrow-band HST imaging. The end of the IR-plume coincides with a warm molecular hydrogen disk and dusty ring, containing 0.7--1.1 $\times$ 10$^7$ M$_{\odot}$  detected within the central kpc. Sensitive upper limits to the detection of cold molecular gas may indicate that a large fraction of the H$_2$ is in a warm state.
Photometry, derived from the UV to the far-IR, shows evidence for a low star formation rate of $\sim$0.04 M$_{\odot}$ yr$^{-1}$ averaged over the last 100 Myrs. A mid-IR component to the Spectral Energy Distribution (SED) contributes $\sim$20$\%$ of the IR luminosity of the galaxy, and is consistent with emission associated with the AGN.  The current measured star formation rate is insufficient to explain NGC3226's global UV-optical "green" colors via the resurgence of star formation in a "red and dead" galaxy. This form of "cold accretion" from a tidal stream would appear to be an inefficient way to rejuvenate early-type galaxies, and may actually inhibit star formation.

 \end{abstract}


\section{Introduction}

A complete picture of the evolution of
galaxies, and how they are grouped into two major branches seen in the local universe, namely the red-sequence and 
the blue cloud \citep{str01,bla03,hog04,fab07} is still a poorly understood problem in astronomy. 
One particular class of galaxy that may be relevant to the discussion
of how stellar populations are built-up over time, are galaxies which
fall in the so-called ``green-valley'', defined in terms of the
UV-optical color lying in the range 3 $<$ NUV-r  $<$ 5, 
a color that places galaxies between the two main color classes. It has been suggested that galaxies in this region 
of color space can be characterized by a flow of star formation--quenched galaxies which migrate from the blue to 
the red \citep{mar07} as star formation shuts down. However, the green-valley galaxies may not be representative of a single 
transitional population \citep{wes07,sch14}, but instead may be complicated by the availability 
of gas supply, and the form of the quenching event that signals the start of a transition from blue to red color.  AGN may play a role
in quenching.  
 
Alternatively, galaxies can potentially move into the green-valley \citep[see][]{thi10}
from the red sequence by a resurgence of star formation through accretion of gas from their environments: either through a gas-rich minor
merger, or through acquisition of tidal debris. 
In this paper we present evidence of a possible on-going HI-gas accretion event in the center
of the elliptical galaxy NGC~3226, and explore its implications for this kind of color change.  
 
The green-valley (NUV-r~=~4.35; this paper) elliptical galaxy NGC 3226 is a companion of the large 
disturbed barred-spiral galaxy NGC 3227, forming the Arp 94 system 
(Figure 1). The core of NGC3226 contains a dusty disk or partial ring (see Figure1-inset), and hints of its LINER-like optical spectra can be traced back to \citet{rub68}. It was spectroscopically classified 
as a LINER- type 1.9 by \citet{ho97}, and is it generally accepted as a classical example of a  low-luminosity, X-ray-bright AGN, most likely the result of radiatively--inefficient accretion onto a supermassive black hole \citep[][ and references therein]{ho01,geo01,ho09a,liu13}. 

Evidence that NGC3226 and NGC3227 are interacting was first presented by \citet{rub68} through optical spectroscopy. 
Subsequent VLA observations revealed 100~kpc--scale plumes of neutral hydrogen extending over 20 arcminutes to the  north and south of the galaxy pair
\citep{mun95}, shown in Figure 2 (left panel), which are most likely tidally--generated structures. 

A contrast-enhanced MegaCam image from Duc et al. (2014), Figure 2 (right panel), exhibits a complex
set of faint stellar arcs and ripples around both NGC 3227 and NGC 3226 \citep[like those described by][]{sch88}. This suggests a rich dynamical history for the system. 
NGC 3226 shows several huge loops and a narrow  optical filament extending from the galaxy to the north-east
at a position-angle of $\sim$30 deg. The visible-light imagery presents a complexity that is hard to reconcile with a single 
tidal interaction between NGC 3227 and NGC 3226. Rather, the structures  around 
NGC 3226 imply that this galaxy is itself the remnant of a recent merger which has launched stellar debris into the joint potential
of what was probably a system of at least three constituent galaxies. 

The HI map shown in Figure 2 seems consistent with two long tidal
tails (north and south) perhaps associated with NGC 3227. However, this may be an oversimplification. Firstly, the system is complicated by the discovery
of a 40 arcsec--scale HI-rich dwarf galaxy (J1023+1952) located in-front of the western disk of NGC3227 \citep{mun95}. This galaxy is physically distinct from NGC3227, shows independent rotation \citep{mun04},  and appears to be experiencing a burst of star formation in part of its disk (see also Figure 1). Studying the molecular and optical/IR properties in more detail, \citet{lis08} suggested that J1023+1952 may be formed at the intersection of two stellar streams.   Secondly, although the southern HI tail contains an optical
counterpart (seen faintly in Figure 2), connecting it to NGC 3227,  the northern plume is less obviously correlated with the faint visible-light structures, and is most likely kinematically associated with NGC 3226 not NGC 3227 (see later).  

We present {\it Spitzer}, {\it Herschel} and {\it Hubble} observations  of NGC 3226, revealing the existence of a warm molecular
disk in its core, which may be being fed from a dusty filament that is likely tidal debris falling-back onto NGC 3226. 
NGC 3226 has a
heliocentric velocity of 1313$\pm$16 km s$^{-1}$ (Simien \& Prugniel
2002), and we adopt a distance of 15.1 Mpc based on a heliocentric
velocity of 1135 km s$^{-1}$ for NGC 3227 (the larger member of the pair) and
H$_0$ = 75 km s$^{-1}$ Mpc$^{-1}$. 

\section{The Observations}

Table 1 summarizes the archival observations used in this paper, including {\it Spitzer} mid-infrared imaging and spectroscopy, 
{\it Herschel} far-infrared photometric mapping, and {\it Hubble}  Space Telescope imaging. 

\subsection{Spitzer Observations}
Some of the mid-IR observations described in this paper where made
shortly after the launch of {\it Spitzer}, during the instrument verification  
phase of the mission.  IRAC \citep{faz04}
imaging in all four bands (3.6, 4.5, 6 and 8~$\mu$m) was performed
using a 7-point dither pattern to encompass both NGC 3227 and NGC
3226, although the emphasis here is on these data for NGC3226.  
Observations using the small-map photometry mode of the MIPS \citep{rie04}
instrument were made at 24$\mu$m.   Infrared Spectrograph \citep[IRS,][]{hou04} photometric mapping 
using the  blue peak-up camera, provided a 
16$\mu$m image of the galaxy as part of the first calibration observations 
of this mode. Raw data from these latter observations were 
processed through the {Spitzer Science Center} standard S18 pipelines, 
and the 16 and 24$\mu$m maps where further processed using the MOPEX mapping package \citep{mak05}. 

A small spectral map was taken in the  IRS Short-Low 
(hereafter SL) module, covering the
wavelength range of 5-15$\mu$m. A 9 $\times$ 1 grid of spectra was obtained
centered on the nucleus of NGC 3226: each single spectrum covered an area  57 $\times$ 3.7 arcsec$^2$.  
 At each slit position, a spectrum 
was taken,  and the slit was moved
2.5 arcseconds (2/3 of a slit width) before taking the next
exposure. Reduction of the spectra were performed using the Spitzer
Science Center (SSC) IRS S18 pipeline, and resulted in a 2-d images of the
spectral-orders.  The spectral cubes were further processed 
using the CUBISM package \citep{smi07a}. Local sky subtraction was performed in the construction of the spectral cubes, using
off-target observations of identical integration time. Spectral line fluxes were extracted using PAHFIT \citep{smi07b}

{\it Spitzer} high resolution (LH = 18.7-37.2 $\mu$m, and Short-High SH = 9.9-19.6$\mu$m modes) spectra taken at the center of NGC 3226 (see Table 1)  were processed through the SSC S18 pipeline.The LH and SH apertures subtend an area of 22.3 $\times$ 11.1 arcsec$^2$ and 11.3 $\times$ 4.7 arcsecs$^2$ respectively.
The two dithered spectra taken in each order were blinked against one another,  to allow identification of cosmic ray glitches and "rogue" pixels, whose values were then replaced by 
interpolation from surrounding pixels. Spectral extractions were made using the SSC software package SPICE ({\it Spitzer} IRS Custom Extraction--a Java based interactive analysis tool), and line fluxes were extracted using the ISAP package \citep{stu98}.

\subsection{Herschel Observations}
{\it Herschel}\footnote{Herschel is an ESA space observatory with science instruments provided by European-led Principal 
Investigator consortia and with important participation from NASA \citep{pil10}}observations at 70 and 160$\mu$m were obtained with the PACS photometer \citep{pog10}, 
and at 250, 350 and 500$\mu$m with the SPIRE (Griffin et al. 2010) photometer.  
Raw data, Level 0, for both instruments were de-archived from the Herschel Science Archive,  and processed to Level 1 using the HIPE (Herschel Interactive processing Environment) 11.1 software (Ott 2006). 
For PACS, map-making beyond Level 1 was performed using the "Scanamorphos" software package (Roussel 2013). These PACS observations combined scan and 
cross scan maps covering 8.3 x 8.3 arcmin$^2$, and were obtained at medium-speed  resulting in a total on-source integration times of  600s.  SPIRE imaging 
\citep{gri10} was performed in large map mode covering an 8 x 8 arcmin area centered on the NGC3226/7 pair. SPIRE data was 
processed through the "Destripper" pipeline in HIPE 11. Details of the observing programs and other information is provided in Table1. 
 
\subsection{Hubble Space Telescope Observations}
{\it Hubble Space Telescope} H$\alpha$ observations (originally discussed by \citet{mar04} in a different context) 
taken with the Advanced 
Camera for Surveys(ACS)  were obtained of the region surrounding NGC 3226 from the {\it Hubble} archive (see Table 1).  
The observations were obtained in the broad-band F814W
continuum filter, 
in which negligible contamination from emission
lines is expected, and in a F658N narrow-band filter,
which includes the redshifted H$\alpha_{\lambda6563}$ and [N{\sc
ii}]$_{\lambda\lambda6548,6584}$ emission lines. Data were processed
with the on-the-fly-calibration procedures on retrieval from the HST archive, and the final
drizzled images were post-processed as follows. The two images were
initially flux calibrated using their respective PHOTFLAM keywords to
ensure correct continuum subtraction with the F814W; the optimal
scaling was verified by ensuring that pure continuum sources, such as
background galaxies, were correctly subtracted. The F814W was shifted
slightly ($\Delta x =-0.46$ pixels, $\Delta y = -0.37$ pixels) for correct alignment
with the F658N image. A revised equivalent PHOTFLAM for the
continuum-subtracted F658N image was then computed using CALCPHOT in
the IRAF SYNPHOT package. Although the F658N bandpass includes
contamination from the [N{\sc ii}] emission lines, their contribution
is assumed to be no more than $\sim$20\% in normal star forming
regions\footnote{If shocks were present this approximation may not be appropriate, and we may overestimate the H$\alpha$ contribution.}.    We therefore followed Martel et al. (2004) and assumed a
single 5$\AA$-wide gaussian line at the H$\alpha$ wavelength appropriate
for the redshift of NGC~3226.  We also present archival V-band
HST observations of the core of NGC 3226 obtained in 1997 with the WFPC2/PC.

\section{Results}

\subsection{A dusty star-forming filament inside the optical dimensions of NGC 3226}

In Figure 3 we show IRAC images of NGC 3226. 
Although this elliptical galaxy shows only a smooth
stellar distribution at 3.5$\mu$m, and 4.5$\mu$m, the galaxy contains a filament of emission
in the 8$\mu$m band (and faintly in the 5.8$\mu$m band). The latter is centered close to the 7.7$\mu$m polycyclic aromatic hydrocarbon (PAH) complex, 
whereas the former may be from the 6.3$\mu$m PAH band.  This structure \citep[noticed by][]{tan09, lan13}, lies well inside
the D$_{25}$ optical dimensions of the galaxy \citep[3.16 x 2.81 arcmin$^2$][]{dev91} and is composed of knots of emission 
extending northwards from the western side of the
inner galaxy. The filament's southern-most extent terminates close to a prominent dust lane seen in optical images of the galaxy (see Figure 1).  We argue
later that the feature is detected in HI emission, and represents the termination of a possible in-falling stream of gas in a tidal tail.

Figure 4 shows the relationship between the IRAC 8$\mu$m emission and the continuum-free HST H$\alpha$+[NII] image, at both full resolution and smoothed to show
extended emission.  HII region complexes are seen associated with the main 8$\mu$m clumps in the filament.  Observed and derived properties from the photometry are presented in Table 2 for the knots labeled A-E in Figure 4. We present the 8$\mu$m flux density and luminosity (Table 2; Column 2 \& 4) converted to a stellar-free flux density S(8)$_{dust}$ by subtracting 25$\%$ of the S(3.6$\mu$m)
flux density from the same aperture from the 8$\mu$m emission to provide a "dust+PAH" band flux \citep{cal07}. The H$\alpha$ flux and luminosity are estimated from the H$\alpha$ HST image for each of the regions (uncorrected for extinction; Table 2; Column 3 \& 5).   

Since the filament was not detected at 24$\mu$m, we estimated the star formation rates
in the filament regions using two methods. The first is based simply on the observed H$\alpha$ luminosity L(H$\alpha$) only \citep{ken98}, (Table 2; Column 5) providing a
SFR(H$\alpha$)(Table 2; Column 7) assuming a Kroupa IMF. This SFR is uncorrected for extinction. A second method, uses both the 8$\mu$m dust(+PAH) luminosity, and the observed H$\alpha$ luminosity from the HST image, and the empirical relationship
of \citet{ken09} between these quantities and the unobscured H$\alpha$$_{corr}$ luminosity (Table 2; Column 6). This  provides an unobscured SFR (Table 2; Column 8). As expected, the
first method yields slightly lower star formation rates than the combined visible/IR method, most likely due to modest extinction in the filament. 
The star formation rates in the filament clumps are very low, between  1-3 $\times$ 10$^{-3}$ M$_{\odot}$ yr$^{-1}$. 

Further insight into the IR properties of the
filament can be obtained by looking at the observations of NGC 3226 at
longer wavelengths.  The filament is detected
at 16$\mu$m, but not 24$\mu$m. Figure 5  shows the close correspondence between the
16$\mu$m emission (contours) and the 8$\mu$m image centered on the 7.7-8.6 PAH bands. Since no
detectable emission is seen at 24$\mu$m (Figure 5, right panel), this implies that the
16$\mu$m image is detecting line emission, either narrow 17$\mu$m 0-0S(1) molecular hydrogen emission, 
or broader 16$\mu$m PAH emission, or both.  It is therefore likely that most of the light in the IR originates in line emission rather than warm dust. 

Emission from PAHs would be expected from the previous discussion on star formation, as the PAHs would be thermally spiked by UV radiation. Since our IRS spectroscopic observations did not cover this region, we cannot conclusively prove whether either large composite planar-bending mode PAHs are responsible for the 16$\mu$m emission, or warm molecular hydrogen. Such emission might be expected if the filament contains shocks.    
 
Figure 6 shows {\it Herschel} PACS and SPIRE images of the galaxy compared with
an optical image on the same scale. The 70$\mu$m map, which has similar spatial resolution to
the {\it Spitzer} 24$\mu$m image of Figure 5 (right panel) shows an unresolved core (on scale $<$5.5 arcsec = 400 pc) with a faint extended (35 arcsec = 2.5 kpc) component elongated roughly along a  position angle of $\sim$ 42-45$\degree$ (N thru E), especially to the north.

The filament may contain colder dust as it is weakly detected in the PACS 160$\mu$m image to the NW of the dusty core, and possibility in the lower resolution SPIRE images (Figure 6).  

\subsubsection{A neutral hydrogen counterpart to the IRAC dust filament}

We show in Figure 7 a series of channel-map images showing HI column density, in steps of 41 km s$^{-1}$, obtained by \citet{mun95} from D-array VLA observations
(contours), superimposed on the IRAC 8$\mu$m image (greyscale) for a
series of correlator channels ranging from a heliocentric velocity
of 1285 - 1120 km s$^{-1}$ . The maps show a clear association
in the centroid of HI emission at each velocity with the 8$\mu$m PAH-emitting
filament. As one proceeds through the channel maps from 1285 to 1120 km$^{-1}$, 
we see the centroid following the filament until the HI emission completely disappears
at velocities lower than 1120 km s$^{-1}$. At higher velocities than 1285 km s$^{-1}$ \citep[not shown here-but see][]{mun04}, the 
centroid joins the bulk of the gas which forms the northern HI filament seen in Figure 2. 
Figure 7 shows two things: i) that the IRAC 8$\mu$m emitting filament is gas
rich (M$_{H}$= 9.2$\times$ 10$^{7}$ M$_{\odot}$ for D = 15.1 Mpc and
integrating over the range of HI channels shown in Figure 7) with the
center of the filament having a systemic velocity of 1202 km/s, and
ii) the star formation seen in this dusty structure seems to represents the southern tip and kinematic termination of the large
northern HI plume. The termination of the northern HI plume deep inside the inner regions of NGC3226 may suggest "late-stage" accretion onto the galaxy.

\subsection{Mid-IR Spectral Properties of NGC 3226}
We show in Figure 8 the IRS-SL spectrum of NGC 3226 extracted
from the spectral cube.  The galaxy shows PAH emission and emission from the 0-0S(3)$\lambda$9.66$\mu$m rotational line of warm molecular hydrogen.  
In Table 3, we present the band and line fluxes for the well determined PAH features and the 0-0 S(3) line determined from the application of PAHFIT \citep{smi07b}. 
 The large value of the 11.3/7.7 (= 0.67) PAH-feature power ratio, and small ratio of the 6.2/7.7 PAH (= 0.22) ratio,  places NGC3226 in the region of neutral PAHs with long carbon chains\citep[N $\sim$300;][]{dra01}. This may relate to a relatively quiescent UV  environment  in NGC 3226 due to its low star formation rate and weak AGN properties.  

In Figure 9 we show the high-resolution spectrum of NGC 3226 obtained with the IRS. Unlike the SL spectrum, a dedicated background observation, off the target, was not made for this archival observation (taken early in the mission). Therefore there is an unknown offset in the spectrum from zodiacal light contamination. However, the line fluxes themselves are not affected. The SH spectrum, extending to 19.5$\mu$m was scaled upwards (in the figure by a factor of 4) so that its continuum matched the overlapping continuum of the LH spectrum which has a larger slit.  Line fluxes are provided in Table 3.  Figure 9 shows strong emission from the rotational H$_2$ transition 0-0S(1)17.03$\mu$m and well as weaker emission
from 0-0S(2)12.28 and S(0)28.22$\mu$m.  Atomic fine-structure lines of [NeII]12.8$\mu$m, [NeIII]15.6$\mu$m,[ SiII]34.81$\mu$m as well as 
PAH emission is also seen in the spectrum.  These observations were taken against a relatively high zodiacal light background (NGC 3226 is only 9 degrees from the ecliptic plane), and therefore the absolute scaling (and possibly the shape) of the continuum is likely affected by extra (zodiacal) background emission. The measurement of the strength of the emission lines is not affected by zodiacal light. 
 
\section{Warm H$_2$ properties}

The discovery of strong pure-rotational emission lines of molecular hydrogen in NGC3226,  implies the existence of warm (T $>$ 100 K) H$_2$ in the galaxy. Temperatures significantly below 100K are not capable of exciting even the lowest level transition which gives rise to the $\lambda$28$\mu$m line. If several emission lines are detected,  it is possible to gain some idea of the temperature and total column density of gas by creating
a H$_2$ excitation diagram: a plot of the upper-level transition column density normalized by its statistical weight N$_u$/g$_u$,  against  the upper-level transition energy E$_u$. If the gas is in thermal equilibrium, the ortho- and para-branches of the H$_2$ energy level diagram contribute points
in the excitation diagram that fall on a straight line, the slope of which is inversely proportional to the temperature \citep[for more details see:][] {rig02,rou07}. 

The inset to Figure 9 shows the excitation diagram derived from the SH/LH spectra, and are based on the line fluxes (Table 3)  for the three lowest energy rotational transitions (0-0 S(0), S(1) and S(2)). We also provide line fluxes for other detected atomic lines in Table 3.  We estimate a range of temperatures and total H$_2$ masses of the warm molecular hydrogen.  Assuming an ortho-to-para H$_2$ ratio
appropriate for LTE, we derive a temperature range of 144--173 K and a warm H$_2$ mass of 1.1--0.7 $\times$ 10$^7$ M$_{\odot}$ in the LH aperture (1.6 $\times$ 0.8 kpc$^2$). The range of temperatures and masses takes into account both observational uncertainty, as well as uncertainty in the relative scaling of the SH and LH spectra which depends on the assumed H$_2$ distribution.

Deep single-dish IRAM 30-m CO observations by \citet{you11} failed to detect molecular hydrogen down to a level of   $<$ 1.7 $\times$ 10$^7$ M$_{\odot}$,
for a nominal velocity width of 300 km s$^{-1}$ and a 3-$\sigma$ upper limit,  and assuming X$_{CO}$= N(H$_2$)/I(CO)$_{1-0}$ = 2 $\times$ 10$^{20}$ cm$^{-2}$ (K km s$^{-1}$)$^{-1}$. Assuming the CO is not very extended on the scale of the 22 arcsec IRAM beam,  and using this X$_{CO}$ factor would imply a  warm-to-total H$_2$ mass ratio, M(H$_2$)$_{w}$/M(H$_2$)$_{total}$ $>$ 0.29-0.39. 

\citet{rou07} investigated the M(H$_2$)$_{w}$/M(H$_2$)$_{total}$ ratio for the SINGs sample, and found that it decreased with
H$_2$ excitation temperature--ranging from 0.5 to 0.02, with a median value of $\sim$ 0.15 (correcting for a slightly different assumed X$_{CO}$ ratio).    It was also found  that the temperature of the warm H$_2$ in LINERs and Seyferts was systematically higher than in HII region-dominated galaxies,  and the M(H$_2$)$_{w}$/M(H$_2$)$_{total}$ ratio was generally lower in these galaxies, as compared with HII-dominated galaxies. 

NGC 3226 lies on the upper end of the distribution of M(H$_2$)$_{w}$/M(H$_2$)$_{total}$ ratios for HII-dominated galaxies, but is an outlier
for LINERs and Seyferts, having more warm H$_2$ than expected for its temperature. The intergalactic filament 
in Stephan's Quintet \citep{gui12} has also been shown to exhibit a high fraction of warm H$_2$, and in that case there is strong evidence that the filament is
strongly heated by mechanical energy from shocks \citep[see also][]{app+13}.  

We caution, however, about over-interpreting the significance of the warm molecular fraction. A comparison of this ratio between galaxies is only meaningful if the value of X$_{CO}$ is the same in NGC 3226 to that in other galaxies, including the Milky-Way \citep[see][for a complete discussion]{bol13}.

To explore the distribution of the H$_2$ in NGC 3226, we created a spectral cube from spectral maps made with the IRS SL data using the CUBISM software of \citet{smi07b}. 
From that cube we extracted a map of the 0-0S(3)9.66$\mu$m  H$_2$ line, after removing a local continuum.  This is shown (black contours) in Figure 10 along with the IRAC 8$\mu$m contours (red contours), both 
superimposed against a B-band image of the galaxy. The emission from H$_2$ is confined to an elongated structure with a PA $\sim$45$\degree$ and on a smaller scale than the PAH emission (which is much more extensive and symmetrically distributed). 

It is reasonable to ask what might be the source of heating of the warm H$_2$ emission? One possibility is that the H$_2$ is heated by the AGN known to lie at the center of NGC 3226. The X-ray luminosity
of the nuclear source \citep{geo01,ho09b} is 4-7 $\times$ 10$^{39}$ ergs s$^{-1}$ (scaled to  D =15.1 Mpc). Summing just the warm H$_2$ line for the 0-0 S(0), S(1) and S(2) lines gives a warm H$_2$ line luminosity in the range 2.0 10$^{39}$ ergs s$^{-1}$ $<$ L(H$_2$) $<$ is  3.4 10$^{39}$ ergs s$^{-1}$, which would imply an impossibly high X-ray gas heating efficiency \citep[the efficiency is expected to be significantly less than 10$\%$][]{lep83}. We can therefore rule out X-ray heating of the H$_2$.  

A common explanation for the detection of emission from the rotational lines of H$_2$ in normal galaxies is through photoelectric heating from small grains and PAH molecules excited in Photo-Dissociation Regions (PDR) associated with young stars. However, for NGC 3226, we estimate the H$_2$/PAH(7.7 $\mu$m ratio) to be 0.06, which places it in the class of galaxies which exceeds the efficiency of H$_2$ heating by PDRs \citep[see][]{rou07} by a large factor (The median value for this ratio for SINGS galaxies is 0.0086). Guillard et al. (2012) used the Meudon PDR code\footnote{See 
http://pdr.obspm.fr/PDRcode.html for details}  to demonstrate that over a wide range of UV excitation and densities in PDRs, the L(H$_2$ S(0)-S(3)) /L(PAH7.7) ratio cannot exceed 0.04. Ratios exceeding this value must have an additional heating source, such as the dissipation of mechanical energy through shocks and turbulence or very high cosmic rays energy densities. 

Evidence that shocks can create very strong heating of H$_2$,  and can boost the far-IR cooling from the [CII]158$\mu$m  line has been presented for the intergalactic filament in the Stephan's Quintet compact group \citep{app06,gui09,clu10,app+13}, the bridge between the "Taffy' galaxies \citep{pet12}, as well as in other Hickson Compact Groups \citep[HCGs;][]{clu13}. \citet{ogl10} defined galaxies with H$_2$/PAH ratios $>$ 0.04 as Molecular Hydrogen Emission Line Galaxies (MOHEGS). Like other MOHEGs, the warm H$_2$ in NGC 3226 is likely heated by mechanical energy through shocks and turbulence. PDR heating must play a minor role, especially given the very low star formation rates we find for this galaxy.

One possibility is that the gas is shock-heated through energy deposited by infall from the HI plume. For a mass infall rate from the plume of  $dm/dt$ = 1 M$_{\odot}$ yr$^{-1}$,   falling from 
a distance of r $\approx$ 10 kpc, and assuming a mass within that radius of $M(r)$ = 10$^{11}$ M$_{\odot}$, the energy input rate would be $dE/dt$ = $(GM(r)/r)~dm/dt$ $\sim$ 3 x 10$^{33}$ W (3 x 10$^{40}$ erg s$^{-1}$), which is enough energy to heat the warm H$_2$ with a 10$\%$ efficiency. 

We cannot completely rule out a role in the shock-heating of the H$_2$ by the AGN. Recently, \citet{ogl14} have shown that even the weak radio jet in NGC 4258 is capable of shock-heating large quantities of molecular hydrogen in that galaxy. NGC 3226 contains a  3.9 mJy flat or inverted spectrum radio source at 5GHz \citep{fil06} which is compact on milli-arcsecond scales \citep{nag05}. However, unlike NGC 4258, there is no direct indication that any radio jet in the system is directly interacting with the host galaxy (in NGC 4258 the so called "anomalous arms" are very obvious at many wavelengths).

\section{The Spectral Energy Distribution of NGC 3226}

\subsection{AGN Contribution to IR}

NGC 3226 is known to contain a low-luminosity AGN, and this might manifest itself in the spectral energy distribution (SED) of the galaxy shown in Figure 11a and b (the flux densities plotted are presented in Table 4). We first consider the infrared component of the SED derived from {\it Spitzer} (IRAC and MIPS 24$\mu$m) and {\it Herschel} far-IR photometry (70 and 120,  250, 350 and 500$\mu$m) derived from a 54 arcsec diameter aperture covering the whole galaxy. Figure 11a shows the observed SED and a model decomposition  of the SED into an AGN and host galaxy component, using the method of Mullaney et al. (2011).  The modeling software developed by these authors,  "DECOMPIR", uses empirical templates for the host galaxy SED, plus an  intrinsic AGN  broken power-law model to fit the SED using a minimization of the least-squares algorithm. In Figure 11a we show the best result from the fitting, in which the AGN fraction of the IR emission, L(AGN)/L(FIR), is 20$\pm$5$\%$. Because of the limited number of empirical host galaxy templates used in this modeling, the final composite spectrum slightly overestimates the 70$\mu$m contribution for the galaxy.  

In order to try to refine the fit, we perform additional modeling using the method of Sajina et al.\  (2006; 2012) that takes into account the full UV-IR SED shown in  Figure 11b.  The additional points are from GALEX (FUV and NUV) sky survey images, SDSS images (u, g, r, i), 2MASS (j, h and k$_s$), as well and the {\it Spitzer} and {\it Herschel} data already discussed.  As with the previous modeling,  a power-law form is assumed for the AGN component.
However, instead of a template for the host galaxy, the galaxy is built up from various model components. These include a stellar photospheric component which assumes a 100 Myr single population model from \citet{mar05}, a warm and cold dust components from Sajina et al.\ (2006), together with the modified PAH template of \citep{lac07}.  Figure 11b presents the least-squares fit of the model to the observed fluxes (black solid line). The red line shows the unattenuated stellar spectrum found to be most consistent with the UV through IR data. Parameters resulting from the model fit are presented in Table 5, including the estimated extinction, star formation rate, dust mass and stellar mass consistent with the SED.  The derived AGN fraction of the far-IR emission is  found to be 18$\pm$4$\%$, very similar to that found using \citet{mul11} model. 

We emphasize that although we have fitted the mid-IR emission with an assumed AGN component, we cannot conclusively prove that the mid-IR component is caused by the AGN. Evidence that the AGN model has some validity comes when we over-plot on Figure 11a (we do not fit) the 16$\mu$m and 24$\mu$m photometry obtained in a small nuclear aperture (diameter 10 arcsec) on the same plot (red points). They agree very closely with the AGN model, suggesting that the mid-IR component is strongly peaked--as would be the case for an AGN.  Previous modeling of NGC 3226 using an earlier version of the SED (without the GALEX or the 16$\mu$m peak-up data) which did not include an AGN has been performed by  \citet{lan13}. 
These authors used a MAGPHYS model \citep{dac08} that assumes all the IR components result form various dusty ISM components. However, as with our modeling, \citet{lan13} found that a hot dust component was needed to explain mid-IR component of the SED, but this model attributes the emission to dust heated in "birth clouds".  Given that this galaxy contains an AGN, we think it more likely that the AGN, rather than birth-clouds, are responsible for the mid-IR power that the SED modeling seems to require.

\subsection{Star formation estimates and other properties based on SED modeling}

The Sajina et al. (2012) modeling described above provides several important parameters that describe the host galaxy, including the star formation rate which is given in  Table 5. The star formation rate  obtained is found to be 0.046 ($\pm$ 0.005) M$_{\odot}$ yr$^{-1}$.  However, in order to provide an alternative estimate, we also repeat the MAGPHYS modeling performed by \citet{lan13}, but using more points.} MAGPHYS constrains a range of physical properties in the galaxy, such as far-IR luminosity, star formation rate, total dust mass and metallicity. This approach has been used successfully in a number of samples, including large SDSS galaxy samples \citep{dac10a}, ULIRGs \citep{dac10b},  and in a large study of early and late type galaxies in HCGs by \citet{bit11}.  We present the results of the fitted parameters in Table 5, along with those measured by \citet{lan13}. Since our values differ only slightly from their fit, we do not present the SED graphically, rather we refer the reader to this paper.   Using this method, we find a total star formation rate of 0.038 ($\pm$ 0.0002) M$_{\odot}$ yr$^{-1}$. These values are similar to those found using the Sajina et al. model. 

The total stellar mass computed by both the Sajina et al, and MAGPHYS models are close to 10$^{10}$ M$_{\odot}$, allowing us to calculate the specific star formation rate  in the range -11.13 $<$ (sSFR) $<$ -11.47  yr$^{-1}$ from both methods. This places it squarely between spiral and elliptical galaxies in the Hickson Compact Group sample of \citet{bit10, bit11}, consistent with its green-valley "transitional" NUV-r = 4.35 color (4.86 before extinction correction). 

A total dust mass was also derived from the MAGPHYS model of  1.24 $\times$ 10$^6$ M$_{\odot}$. To confirm that this mass is reasonable, we also estimate the dust mass using the formulae of both \citet{mag12,mag13} as well as  \citet{dun11}. Assuming T~15K and $\beta$ =1.5 we 
obtain 1.67 $\times$ 10$^6$ M$_{\odot}$ and 1.64 $\times$10$^6$ M$_{\odot}$ respectively. Changing the assumed value of $\beta$ from 1.5 to 2, reduces the dust mass by 5$\%$.  These values are within 25-30$\%$ of the MAGPHYS results, and provide some measure of the uncertainty inherent in the different assumptions used to estimate the dust mass.  

If the dust mass measured by these methods is accurate, then we can estimate the total gas mass for an assumed dust to gas (DGR) ratio. As shown by \citet{san13}, the DGR is strongly dependent on metalicity. If we assume DGR of 0.16, a value appropriate for Milky-way metalicity galaxies, this would predict 8 $\times$ 10$^{7}$ M$_{\odot}$ of gas in the system, more than four times the measured upper limit to the molecular gas from the IRAM CO observations ($<$ 2 x 10$^{7}$ M$_{\odot}$).  On the other hand, it is well known that  the conversion of I(CO) to N(H$_2$), X$_{CO}$, need not be constant throughout the universe.  One possibility is that X$_{CO}$ is higher than that assumed, perhaps because the gas acquired by NGC3226 was of lower metallicity.  There is some evidence that the X$_{CO}$ factor rises rapidly in gas with  metalicities below 1/3-1/2 solar \citep[see][]{bol13}.

\section{The recent evolution of NGC 3226}

NGC 3226 is far from a normal elliptical galaxy. Although there is evidence that it is strongly interacting with NGC 3227,
the complex web of optical and HI gas in the Arp 94 system as a whole suggests that this is not a simple two-body interaction.
The optical ripples around both galaxies (see Figure 1), and the complex loops and optical spurs around NGC 3226 may suggest that this galaxy itself is the result
of a recent merger, the tidal debris of which may have become shared with NGC 3227.   We concentrate here only on the potential connection between
northern HI plume, and recent star formation 
in and around NGC 3226, as a complete understanding of the many complexities of Arp 94 is beyond the scope of this paper. 

There are several possibilities for the origin of the IRAC-detected filament. The first
is that it is a structure tidally spun out from NGC 3226 from some
pre-existing ring of material orbiting the galaxy--perhaps originally
captured from the gas-rich companion at an earlier time, or, more
likely, it is the result of late-stage accretion onto NGC 3226 from
the large neutral hydrogen plume. We favor the second explanation on
kinematic grounds, and because material falling back onto
both galaxies is expected in mature tidal systems like Arp 94 from
modeling arguments \citep{str97}. 

{Asymmetric strong dust absorption structures, such as the 
incomplete dust absorption ring structure is seen in the deep HST B-band image of NGC 3226, and reproduced in Figure 12, have often been associated with possible infall events \citep{tra01,mar04}.} We superimpose the warm H$_2$ distribution from the IRS cube as well as a schematic of the position of
the IRAC filament on the image.  One possibility is that the dusty/PAH emitting filament (which is associated with the HI plume)
is feeding the warm H$_2$ disk and dust-lanes seen in Figure 12. As discussed by \citep{sim07}, the presence of nuclear dust lanes may also correlate with AGN activity. 

Is there a kinematic connection that might link the structures? 
We  note that the systemic velocity of NGC 3226 (determined optically) is
1313 ($\pm$13) km s$^{-1}$, approximately 100 km s$^{-1}$ higher than the central
velocity of the IRAC filament (measured using HI).  Atlas$^{\rm3D}$ IFU data \citep{kra11} provides information on the rotation of the stars 
(91 km/s peak values along a kinematic major axis PA 28 of deg.).     
The scale of this rotating stellar core is large enough to extend
towards the point where the IR filament reaches the minor axis.  More recently unpublished Atlas$^{\rm3D}$ 
imaging of the ionized gas in NGC 3226 (M. Sarzi, University of Hertfordshire;  Private Communication) 
shows a clear connection between a rotating ionized disk in NGC3226 and the filament--with 
ionized gas in the filament having a similar velocity to the HI (around 1260-1280 km s$^{-1}$).  
It is not unreasonable to speculate that the HI/dust filament may be feeding
both the warm H$_2$ disk and the ionized disk. 

As  \citet{dav11}  have pointed out, in fast rotators like NGC 3226, there are many cases where aligned (or almost aligned) stellar  and ionized gas disks exist in galaxies that have clear evidence of external accretion. In this case the kinematic axes of the stellar and ionized-gas disks are not strongly divergent, with position angles of 
28$\degree$ and 36$\degree$ respectively \citep[see also][]{kra11} with an uncertain of 6$\degree$. We measure the position angle of the major axis of the incomplete dust ring in Figure 12 to be very similar (28$\degree$) to the stellar disk rotation major axis, suggesting that the dust has settled into the rotational plane of the stars. However, we find that the position angle of the warm molecular disk  45$\degree$, is significantly different from the other components, indicating that it may be dynamically unrelaxed. 
 
A rough estimate of the accretion rate onto NGC 3226 can be
obtained by dividing the HI mass in the dusty part of the filament by
the fall-back time onto the galaxy. There is a small velocity gradient
along the filament (from inspection of Figure 6 of Mundell et
al. 1995) which is 2.9 km s$^{-1}$ kpc$^{-1}$. Inverting this gives a timescale of
approximately 300 Myr, which would be an upper limit to
the time taken for gas to move along the filament. This timescale could be much shorter since the free-fall time for gas at 10 kpc for a galaxy of mass 10$^{11}$ M$_{\odot}$ is approximately 70 Myr. However, given the effects of tidal interactions, gas is unlikely to fall radially onto the galaxy, and so the actual infall rate may be somewhere between these two values.  Taking the HI mass
of the  inner filament (we only include the gas shown in Figure 7 which is contained within 5 kpc of the galaxy) we obtain a very approximate accretion rate of dM/dt $>$
0.3 M$\odot$ yr$^{-1}$. As discussed above this is likely a conservative lower limit because the infall time could shorter.

This rate is close that needed to heat the warm H$_2$-see Section 4. This accretion rate implies that 
10$^8$ M$\odot$ of HI could pile up into the center of NGC 3226 over a few hundred Myrs. Hence
a substantial disk of material could be formed in the core of the
galaxy during the later stages of the collision. We have already shown
that the warm molecular hydrogen disk has an H$_2$ mass of 0.75--1.1 $\times$ 10$^{7}~$M$_{\odot}$, which could be
supplied to the galaxy through from the HI filament. Since the inner parts of NGC 3226 contain both ionized and
warm molecular gas, it is possible that the mainly neutral in-falling gas is shock-heated as it falls into the nucleus. Attempting to model this possible
accretion process through various ISM phases is beyond the scope of this observational paper.

\subsection{Is the measured SF activity consistent with the galaxy transitioning from red to green by resurgence of star formation?}

The very low star formation rate obtained from the UV to far-IR SED fitting, along with the predominantly warm state of the molecular hydrogen
in the core, argues against the idea that
NGC3226 could lie in the UV-optical green-valley because of a possible resurgence of star formation. For example, a reasonable "extreme" limit for
a galaxy to get from a red state to a green-valley state is if 5$\%$ of its mass was experiencing a star burst (Schawinski et al. 2009). 
Smaller ratios of young to old stellar mass would fail completely to move a galaxy away from the (optical) red--sequence (Schawinski et al. 2007, Kaviraj et al. 2007).
Based on our SED modeling we estimate the total stellar mass of NGC 3226 to be 1.1 $\times$ 10$^{10}$ M$_{\odot}$, and so to generate 5$\%$ of this 
mass with a star formation rate of 0.04 M$_{\odot}$ yr$^{-1}$ would require approximately a Hubble time. Therefore, at its present rate, it is quite implausible that
the star formation rate we presently see could be responsible for the green-valley colors of the galaxy. Of course, the SFR of NGC 3226 may 
have been much larger in the past. In order to significantly perturb normal elliptical colors, at least 5-10$\%$ of the mass, or 5-10 $\times$ 10$^{8}$ M$_{\odot}$ 
would be required. Even if all the warm H$_2$ that we measure (0.7--1.1 x 10$^7$ M$_{\odot}$) plus any possible unobserved cold gas (based on a good upper limit 
from IRAM) were to turn into stars rapidly, this would not be enough to generate enough blue stars to push the galaxy into the green-valley. Thus the green-valley 
nature of NGC 3226 must result from some other cause. 

\subsection{Does NGC 3226 fit into a broader picture of  star formation "quenched" galaxies?}

It is tempting to see NGC 3226 as a pathological case which shares little in common with other green valley galaxies.  However, 
the environment of NGC 3226 may share some similarities with denser galaxy environments, like compact groups.
This would be especially true if, as we suspect, NGC 3226 is actually itself a relatively recent merger product . The loops and filaments
spread throughout the system shown in Figure 2 certainly suggest that the Arp 94 system was composed of more than two galaxies in the past. 

Recently Cluver et al. (2013) has shown that 10$\%$ of HCG galaxies studied by the {\it Spitzer's } IRS contain galaxies with enhanced L(H$_2$)/L(PAH$_{7.7}$) ratios. These MOHEG-classified HCG galaxies predominantly lie in the UV-optical green valley, and have unusually low sSFR.  NGC 3226 shares similar properties to these galaxies.  Johnson et al (2007) and  Walker et al. (2010) suggested that these galaxies are undergoing some kind of transition from the blue cloud to the green valley based on their IR colors, and  Cluver et al. (2013) suggested that shocks and turbulence may partly be responsible for their green-valley colors by helping to suppress star formation. In one example studied in detail, HCG57A, \citet{ala14a} have shown that star formation seems to be suppressed in regions of the galaxy which are experiencing large disturbances in the CO-measured velocity-field caused by a collision with a companion. NGC 3226, which is clearly disturbed, shares some of the same properties of these unusual HCG galaxies. 

If NGC 3226 is really a composite galaxy formed by the recent merger of two smaller galaxies, then perhaps the green-valley nature of NGC 3226 can be explained as a galaxy caught in the late stages of star formation quenching.  For example, the merger may have largely exhausted its gas supply,  and its ability to form many new stars. Further suppression by the action of shocks from external tidal accretion may also play a role. Some support for this picture can be seen by considering the WISE  \citep[Wide-field Infrared Survey Explorer;][]{wri10} colors of the galaxy.  Figure 13 shows a WISE color-mass diagram from \citet{ala14b} showing a strong bifurcation of galaxies into late-type and early type galaxies from a spectroscopically selected SDSS sample.  Alatalo et al. show that there exists an Infrared Transition Zone (IRTZ) which likely contains quenched galaxies which are moving from the top part of the diagram (blue contours = gas rich late-type galaxies) towards the lower part of the diagram (red contours = red cloud galaxies). The IRTZ galaxies are shown to predominantly have the spectral signatures of LINER galaxies and shocked post-starburst galaxy candidates based on their optical spectral-line ratios.   NGC 3226 falls within the lower part of the transition zone\footnote{We obtained the WISE 4.6 and 12$\mu$m magnitudes from the ALLWISE  survey catalog \citep{wri10}. NGC3226 has w2gmag (4.6$\mu$m) and w3gman (12$\mu$m) values of 8.666$\pm$0.008 and 7.25$\pm$0.02 mag.}.  It is therefore consistent with being part of a larger transitional population of galaxies in which shocks or low-luminosity AGNs play a role in transitioning the galaxies.  Further work on the stellar populations in NGC3226, such as searching for evidence of a post-starburst population hinting at recent quenching, would be worthwhile.  

\section{Conclusions}

By combining mainly archival infrared observations from {\it Spitzer} and {\it Herschel}, with optical
images from HST we have expanded our understanding of the green-valley elliptical galaxy
NGC~3226 and have come to the following conclusions:

1) A narrow filament is detected in the mid-IR extending into the core of the galaxy, approximately
terminating at the scale of a dusty ring or partial spiral seen in deep optical HST images and a warm H$_2$ disk (next item). 
The filament is closely associated with HI emission from a much larger HI plume which extends to the north, but
which kinematically ends at the filament, and is spatially coincident with it. The filament--which mainly glows in the light
of PAH molecules and probably molecular hydrogen emission, is likely heated by string of tiny HII regions (SFR $\sim$3 x 10$^{-3}$ yr$^{-1}$) 
which we detect at the termination of this HI plume. The median star formation rate in the filament clumps is 
about 7$\%$ of the star formation rate of the entire galaxy. We suggest that the HI plume is feeding material into the center of the galaxy. 

2) Spitzer IRS observations of NGC 3226 show the existence of a 1 kpc-scale warm disk of
molecular hydrogen of 0.75--1.1 $\times$ 10$^{7}$ M$_{\odot}$ which  has a different orientation (by 17$\degree$) from the major-axis
of the dusty incomplete nuclear ring, and the position angle of the major axis of the fast stellar rotation seen in the galaxy's core.
This differences may imply that the warm molecular disk is not yet in dynamical equilibrium with the other components.

3) We can rule out heating of the warm H$_2$ gas disk by either X-rays from the low-luminosity AGN, or a dominant PDR component associated with
star formation. Instead, we suggest that the H$_2$ line luminosity can be explained by shock heating.  We favor HI gas accretion 
as the source of the mechanical heating.  An accretion rate of 1 M$_{\odot}$ yr$^{-1}$ would be needed (assuming a 10$\%$ efficiency) to balance the observed warm H$_2$ line luminosity, which we show is plausible. 

4) The bulk of the molecular gas in NGC 3226
is warm (with M(H$_2$)$_{warm}$/M(H$_2$)$_{cold}$ $>$ 0.3) based on an upper limit to the detection of CO in the system. Such gas may not be conducive to significant star formation if it is turbulently heated 
\citep[see][]{clu10}. Alternatively, the conversion factor X$_{CO}$ assumed for Galactic emission may be a factor of 4 higher in NGC 3226 if the in-falling gas has low metalicity.    

5) We measure the star formation rate in the galaxy globally to be
very small ($\sim$ 0.04 M$_{\odot}$ yr$^{-1}$). Since NGC~3226 lies in the UV-optical green-valley ([NUV-r]$_{corr}$ = 4.35), we show that
the current star formation rate in NGC 3226 is insufficient to explain the "green" colors of the galaxy based on a resurgence of
star formation, unless the star formation rate was significantly higher in the past. The SED is also consistent with the existence
of a hot dust component which we associate with the AGN, which contributes no more than 20$\%$ to the bolometric luminosity of the galaxy.

6) NGC 3226 has some similar properties to a subset of early-type galaxies in Hickson Compact Groups that may be undergoing rapid evolution (low specific star formation rate, green UV-optical colors, large warm H$_2$/PAH ratio). We also show that NGC 3226
lies in the newly discovered WISE infrared transition zone for SDSS galaxies \citep{ala14b}. This zone, which may signify evolution from dusty late-type galaxies to dust-free early type galaxies, has been shown to be dominated by galaxies with LINER and post-starburst shocked-gas optical spectra. NGC 3226 may, despite its complex dynamical environment, fall on a continuum of galaxies undergoing star formation quenching.

\acknowledgements
We wish to thank an anonymous referee for very helpful suggestions on how to improve the paper. This work is based, in part, on observations (and archival observations) made with the {\it Spitzer}  Space Telescope, which is operated by the Jet Propulsion Laboratory, California Institute of Technology under a contract with NASA. The work is also based, in part, on observations made with {\it Herschel}, a European Space Agency Cornerstone Mission with significant participation by NASA. Partial support for this work was provided by NASA through an award issued by JPL/Caltech. TB \& VC would like to acknowledge partial support from EU FP7 Grant PIRSES-GA-20120-316788. CGM acknowledges support from the Wolfson Foundation, the Royal Society and the Science and Technology Facilities Council. UL acknowledges support by the research projects   AYA2011-24728 from the
Spanish Ministerio de Ciencia y Educaci\'on and the Junta de Andaluc\'\i a (Spain) grants FQM108. PA would like to thank Marc Sarzi,  and  Tim de Zeeuw for information about the ionized gas in NGC 3226 based on unpublished ATLAS$^{\rm3D}$ data, and Sergio Fajado-Acosta for advice on the calibration of the IRS peak-up mode.

\newpage
\begin{deluxetable}{lrrrr}
\tabletypesize{\scriptsize}
\tablecaption{Journal of {\it Spitzer} and {\it Herschel} Observations}
\tablewidth{0pt}

\startdata
\tableline 
\tableline \\
Instrument &UT Date&Instrument mode & Integration time & Wavelength($\mu$m) \\
                   &        &          &  on source (s)          &                                   \\
\tableline \\ 
{\it Spitzer} IRAC$^a$ & 11:23:2003 &7-pt. med. dither & 288  & 3.6,4.5,6.8,8.0 \\
{\it Spitzer} MIPS$^a$ & 11:24:2003 &Small-map photomety & 300 & 24 \\
{\it Spitzer} IRS$^b$ & 12:17:2003 &Peak-up imaging map & 60 & 16 \\
{\it Spitzer} IRS$^c$ & 11:28:2003 &Short-low 9 x 1 map & 60 & 5-14 \\
{\it Spitzer} IRS$^d$ & 05:26:2005 &Short-Hi/Long-Hi staring &  365/487 & 9.7-37.2 \\
{\it Herschel}$^e$ PACS & 15:16:2011 &Photometry Scan Map & 600 & 70 and 160 \\
{\it Herschel}$^f$ SPIRE &  05:30:2010 &Photometry Large map & 1112 & 250, 350  and 500 \\
{\it Hubble}$^g$ ACS & 03:08:2003  &F814W filter & 700 & 0.8060, 0.6583.9 \\
{\it Hubble}$^h$ WFPC2/PC & 03:18:1997 &F547M filter & 460 & 0.547 \\
\enddata
\tablenotetext{a}{Observations from P. N. Appleton; Program ID = 1054}   
\tablenotetext{b}{"In-orbit check-out" observations from J. Houck; Program ID = 668} 
\tablenotetext{c}{"In-orbit check-out" observations from J. Houck; Program ID = 1401}
\tablenotetext{d}{Archival observations from C. Leitherer; Program ID = 3674}
\tablenotetext{e}{Archival observations from  GT1$\_$lspinogl$\_$2; obsids = 1342221146}
\tablenotetext{f}{Archival observation from  KPOT$\_$seales01$\_$1; obsid =1342197318}
\tablenotetext{g}{Archival observations from GTO/ACS program 9293; PI. H. Ford} 
\tablenotetext{h}{Archival observation from WFPC2/PC program 6837; PI. L. Ho}
\end{deluxetable}

\newpage

\begin{deluxetable}{lllllll}
\tabletypesize{\scriptsize}
\tablecaption{Observed and Derived IR Properties of Knots in NGC 3226 Filament}
\setlength{\tabcolsep}{0.01in}
\startdata
\tableline
\tableline \\
Knot & S(8$\mu$m)$_{dust}$$^{a,b}$ & H$\alpha$$_{unc}$$^{b,d}$ flux (x 10$^{-15}$ &Log(L(8$\mu$m))$^b$ & log(L(H${\alpha}$$_{unc}$))$^{b,d}$ & SFR(H${\alpha}$$_{unc}$))$^{b,c}$ &  SFR(8$\mu$m+H${\alpha}$)$^{b,e}$\\
         &   (mJy)      &  ergs s$^{-1}$ cm$^{-2}$)  & (erg s$^{-1}$)  & (erg s$^{-1}$) & (10$^{-3}$ M$_{\odot}$yr$^{-1}$) & (10$^{-3}$ M$_{\odot}$yr$^{-1}$)\\          
(1) & (2) & (3) & (4) & (5) & (6) & (7) \\
\tableline
A & 0.51 (0.1)& 1.5 (0.15) & 39.71 &  38.23 & 0.9 &  1.2  \\
B & 0.42 (0.1) & 4.7 (0.2) & 39.63 &  38.73 & 2.8 & 3.1 \\
C & 0.91 (0.1)& 5.0 (0.2)  & 39.96 &  38.76 & 3.0 & 3.6 \\
D & 0.82 (0.1)&  4.6 (0.2)  & 39.92 & 38.72 & 2.8 & 3.3 \\
E & 0.53 (0.1)& 4.4  (0.2) & 39.81 & 38.70 & 2.7  & 3.0 \\
\enddata
\tablenotetext{a}{8$\mu$m flux density corrected by factor of 1.2 for r = 3.6 arcsec aperture correction(Table. 5.7 of the IRAC Data Handbook). The flux was also corrected to a "dust+PAH" flux by subtracting 0.25 x S(3.6)$\mu$m flux density to remove
the contribution to the band from stellar photospheric emission in the galaxy \citep[see][]{cal07} }
\tablenotetext{b}{evaluated over aperture radius 3.6 arcsecs, or area = 0.24 kpc$^2$ at D= 15.1 Mpc} 
\tablenotetext{c}{Assuming SFR M$_{\odot}$ yr$^{-1}$ = 5.3 x 10$^{-42}$ L(H$\alpha$) and Kroupa IMF \citep{ken98} } 
\tablenotetext{d}{H$\alpha$ fluxes (or luminosity) uncorrected for extinction derived from from the HST image.}
\tablenotetext{e}{SFR calculated using the empirical relationship between Log(L(H$\alpha$$_{unc}$)+0.011L(8$\mu$m)) and Log(H$\alpha$$_{corr}$) \citep{ken09}, where H$\alpha$$_{unc}$ is uncorrected for 
extinction, and H$\alpha$$_{corr}$ is the unextinguished values, and then converted to SFR using table-note (c).} 
\end{deluxetable}

\begin{deluxetable}{llllllll}
\setlength{\tabcolsep}{0.01in}
\tabletypesize{\scriptsize}
\tablecaption{Extracted Line and PAH Band Fluxes from {\it Spitzer} IRS (SL, SH and LH)  in units 10$^{-17}$W m$^{-2}$}

\startdata
\tableline
\tableline \\
IRS Module & 6.2$\mu$m PAH & 7.7$\mu$m PAH & 11.3$\mu$m PAH & 12.6$\mu$mPAH  & 0-0S(3)9.7$\mu$m & 11.3/7.7 & 6.2/7.7 \\
\tableline \\
SL$^a$  & 15.0 (1.5) & 67.1 (6.7) & 44.6 (4.5) & 19.9 (2.0) & 2.35 (0.3) & 0.66 & 0.22\\                  
\\
\tableline
\tableline \\
IRS Module & 0-0S(0)28.2$\mu$m & 0-0S(1)17.0$\mu$m & 0-0S(2)12.3$\mu$m & 11.3$\mu$m PAH &  [NeII]12.8$\mu$m & [NeIII]15.5$\mu$m & [SiII]34.8$\mu$m \\
\tableline \\
LH  &   1.8 (0.15) & - & - & - & - & - &  2.53 (0.29) \\
SH$^b$      & - & 9.2 (0.4) &  1.5 (0.6)   &  94 (10) & 7.8 (0.67) & 5.7 (0.24) & - \\
SH$^c$(Preferred) & - & 4.6 (0.2) &  0.75 (0.3) &  47 (6)   & 4.0 (0.34) & 2.8 (0.12) & -  \\
\enddata
\tablenotetext{a}{Integrated over an area of 22 x 16 arcsec centered on the H$_2$ disk and capturing most of its emission. Line fluxes were measured using PAHFIT \citep{smi07b}.}
\tablenotetext{b}{Assumes the H$_2$ is extended on the scale of the LH slit (22 x 11 arcsecs$^2$) and SH scaled by 4.1 to LH}
\tablenotetext{c}{Assumes the H$_2$ is extended on a more realistic intermediate scale (approx. 15 x 15 arcsecs$^2$), and SH scaled by 2.0 to LH. Supported by similarity in PAH 11.3$\mu$m flux as measured in SL and SH, as well as the measured scale of the H$_2$ seen in Figure 10. } 

\end{deluxetable}

\begin{deluxetable}{llllllll}
\tabletypesize{\scriptsize}
\tablecaption{Observed Flux Density of NGC 3226 from UV to Far-IR and 1$\sigma$ uncertainties--54 arcsec aperture}

\startdata
\tableline
\tableline \\
UVW2$^a$ &  NUV$^b$  & UVW1$^a$ & SDSS$_u$$^c$ & SDSS$_g$$^c$ & SDSS$_r$$^c$ & SDSS$_i$$^c$ & SDSS$_z$$^c$  \\
(mJy) &   (mJy) & (mJy) & (mJy) & (mJy) & (mJy) & (mJy) & (mJy) \\
\tableline \\
0.86 (0.09) &  1.24 (0.12) & 2.65 (0.39) & 8.71 (0.45) & 49.2 (2.0) & 110.7 (6.0) & 166 (9) & 211 (10) 
\\
\tableline 
\tableline \\
J$^d$ & H$^d$ & K$_s$$^d$ & IRAC-3.6$\mu$m$^e$ & IRAC-4.5$\mu$m$^e$  & IRAC-5.8$\mu$m$^e$  & IRAC-8$\mu$m$^e$  &  IRS-16$\mu$m$^e$ \\  
(mJy) & (mJy) & (mJy) & (mJy) & (mJy) & (mJy) & (mJy) & (mJy) \\
 \tableline \\
275 (25) & 286 (28) & 249 (25) &117 (7) & 74 (4) & 51.9 (2.8) & 44.0 (1.4) & 32 (6) \\
\\
\tableline 
\tableline \\
MIPS-24$\mu$m$^e$  & PACS-70$\mu$m$^e$  & PACS-160$\mu$m$^e$  & SPIRE-250$\mu$m$^e$  & SPIRE-350$\mu$m$^e$  & SPIRE-500$\mu$m$^e$ & & \\
(mJy) & (mJy) & (mJy) & (mJy) & (mJy) & (mJy) & & \\
\tableline \\
67 (7) & 332 (17) & 1133 (56) & 722 (4) & 367 (15) & 123 (10) & &\\

\enddata
\tablenotetext{a}{From the SWIFT satellite survey.}
\tablenotetext{b}{Derived from image from the GALEX sky survey.}
\tablenotetext{c}{Data obtained from the SLOAN Digital Sky Survey.} 
\tablenotetext{d}{Data from 2MASS. }
\tablenotetext{e}{Data from this paper.}

\end{deluxetable}

\begin{deluxetable}{llllllllll}
\tabletypesize{\tiny}
\setlength{\tabcolsep}{0.02in}
\tablecaption{Derived Properties of NGC 3226 from SED Fitting}
\startdata
\tableline \\
\tableline \\
Source & $\tau$$_v$ & Log(sSFR) & Log(SFR) &  Log(L$_{AGN}$) &  Log(L$_{dust}$) & Log(Stellar Mass) & Log(M$_{dust}$) & Cold Dust & Warm Dust \\
           &                    &  (yr$^{-1}$)  & (M$_{\odot}$ yr$^{-1}$) & (L$_{\odot}$) & (L$_{\odot}$) &  (M$_{\odot}$) & (M$_{\odot}$) & Temp. (K) & Temp. (K) \\
\tableline
AGN+SF$^a$          &  1.93 &   -11.13   &  -1.34$^{f}$    & 7.70 &  --  & 9.79  & 6.62  & 16.34         & -- $^a$   \\       
MAGPHYS$^b$ & 0.34$^c$& -11.47  & -1.43  & -- &8.63  &10.07  &  6.09  & 17.2 & 59.4   \\
\citet{lan13}$^d$   & 0.91   &  -11.67$^e$ & -1.35 &  & 9.14 &10.36 & 6.00 & 24.1 & 54.4 \\
\enddata
\tablenotetext{a}{From Sajina et al.\ (2006). The model does not assume a specific temperature for the warm dust component} 
\tablenotetext{b}{Differs from Lanz et al. (2013) by the addition of GALEX and 16$\mu$m IRS peak-up point} 
\tablenotetext{c}{We assume A$_v$ = 1.086 x $\tau$$_v$ and find A$_v$  = 0.37}
\tablenotetext{d}{Also uses MAGPHYS with a subset of the points used above} 
\tablenotetext{e}{Correcting a transcription error in Lanz et al. (2013) for sSFR}
\tablenotetext{f}{The SFR was estimated by combining the cold and warm dust emission components to form infrared luminosity $L_{FIR}$ = 2.64$\times$ 10$^8$, and then converting to SFR assuming ${\rm log_{10}}(SFR)={\rm log_{10}}(L_{FIR})-9.762$.}

\end{deluxetable}

\newpage
\begin{figure}[!ht]
\vspace{0.3in} 
\plotone{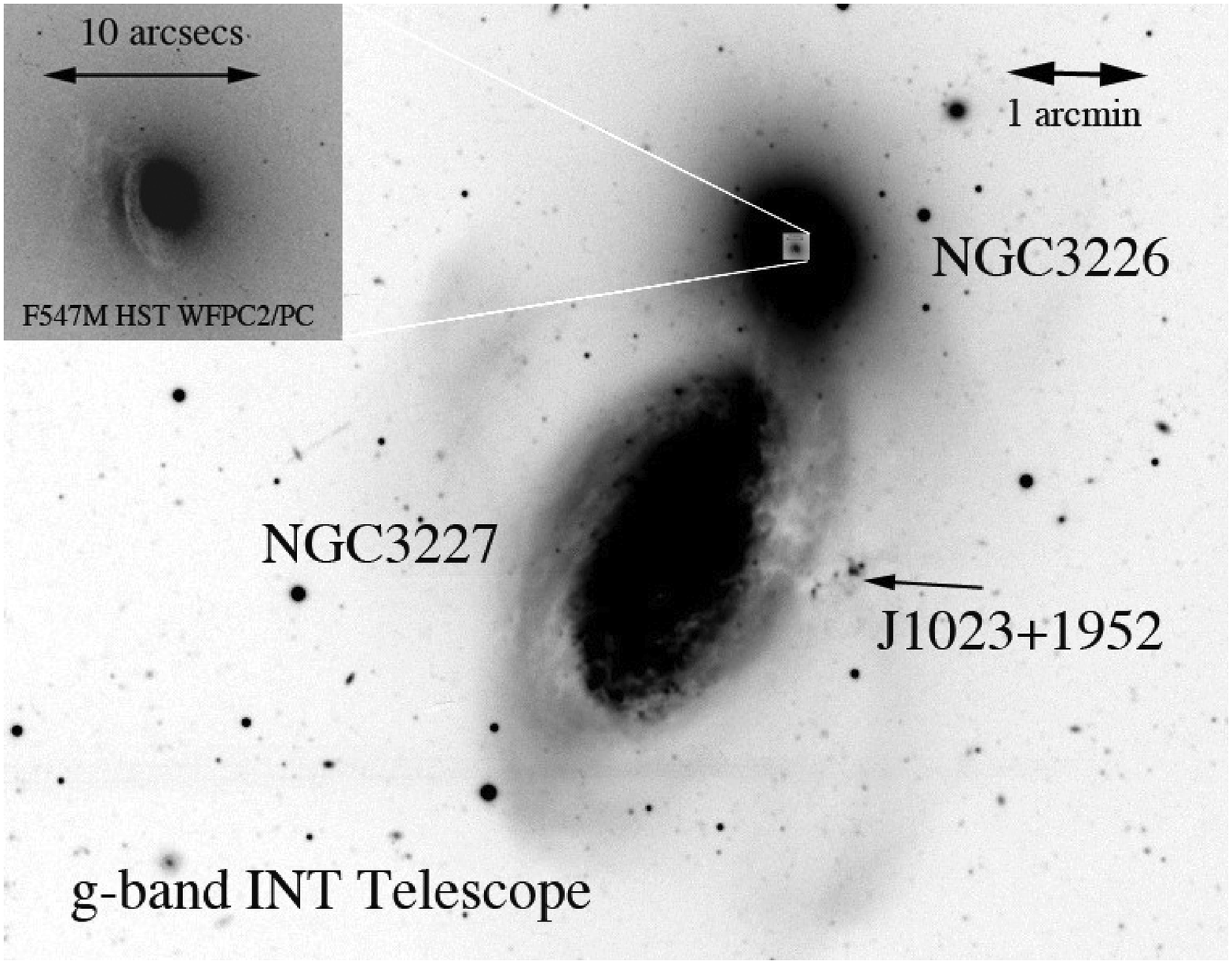}
\caption{g-band image of the Arp 94 system. NGC 3226 lies to the north of NGC 3227 and the HI-rich dwarf galaxy J1023+1952 to the west.
The inset shows a V-band HST image of the core of NGC3226 which contains a partial dust ring or disk. 
}
\end{figure}

\newpage
\begin{figure}[!ht]
\centering
\includegraphics[width=3in]{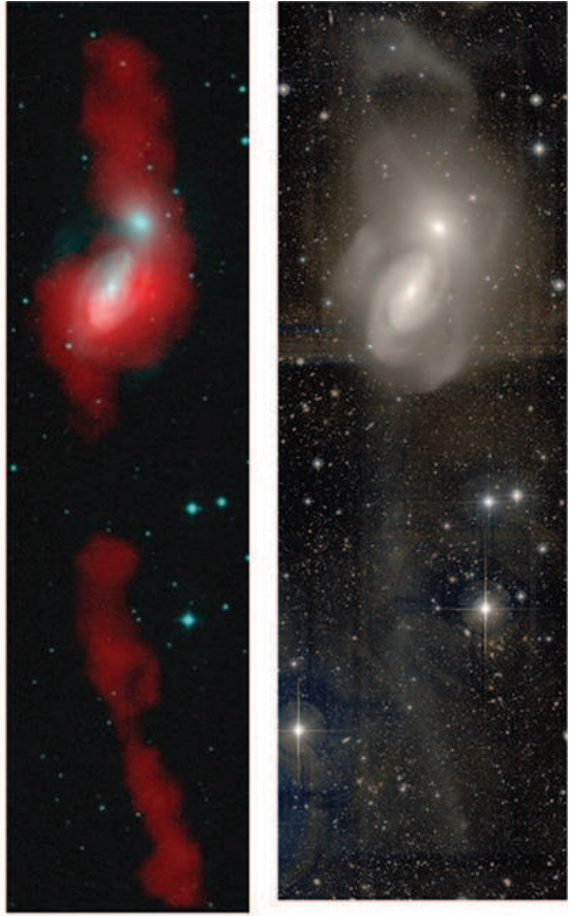}
 \caption{Left: The large-scale neutral-hydrogen emission (red) superimposed on a deep optical image of the galaxy, from observation by Mundell et al. (1994). The figure shows the large scale ($\sim$ 20 arc minutes) of the HI plumes, suggesting that the members of Arp 94 have interacted tidally in the past. Right: Composite g+r MegaCam/CFHT image of the Arp94 system (Duc et al., 2014) rendered with an arcsinh stretch to bring out the faint structure. The image, matched in scale to the HI image, shows evidence of considerable tidal debris, including shells and ripples. The southern HI plume appears to have a faint stellar counterpart, whereas the northern filament is less correlated with the faint optical light. }
\end{figure}

\begin{figure}[!ht]
\vspace{0.3 in} 
\plotone{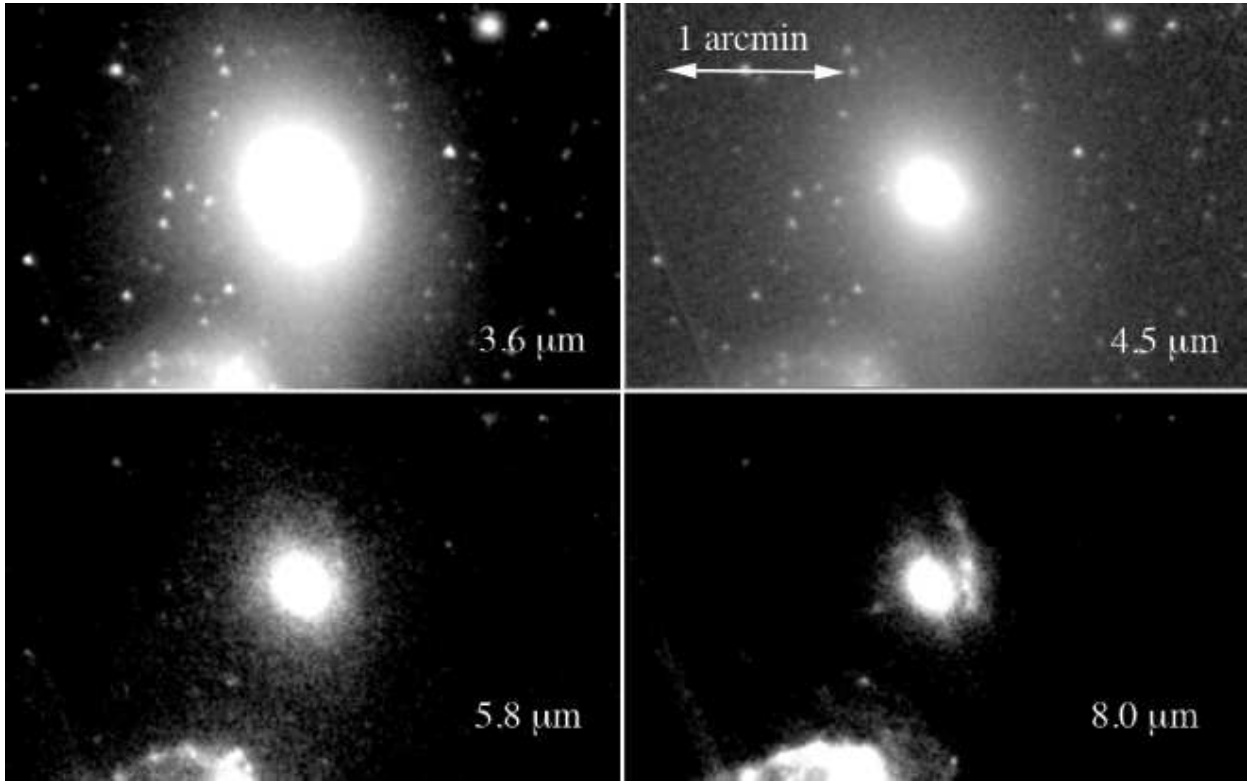}
\caption{A greyscale representation of the 3.6, 4.5, 5.8 and 8.0$\mu$m {\it Spitzer} IRAC images of NGC3226. North is to the top and east is to the left. Note the filament
seen embedded within the galaxy in the 8$\mu$m image. and more faintly at 5.8$\mu$m. }
\end{figure}

\begin{figure}[!ht]
\vspace{0.3 in}
\plotone{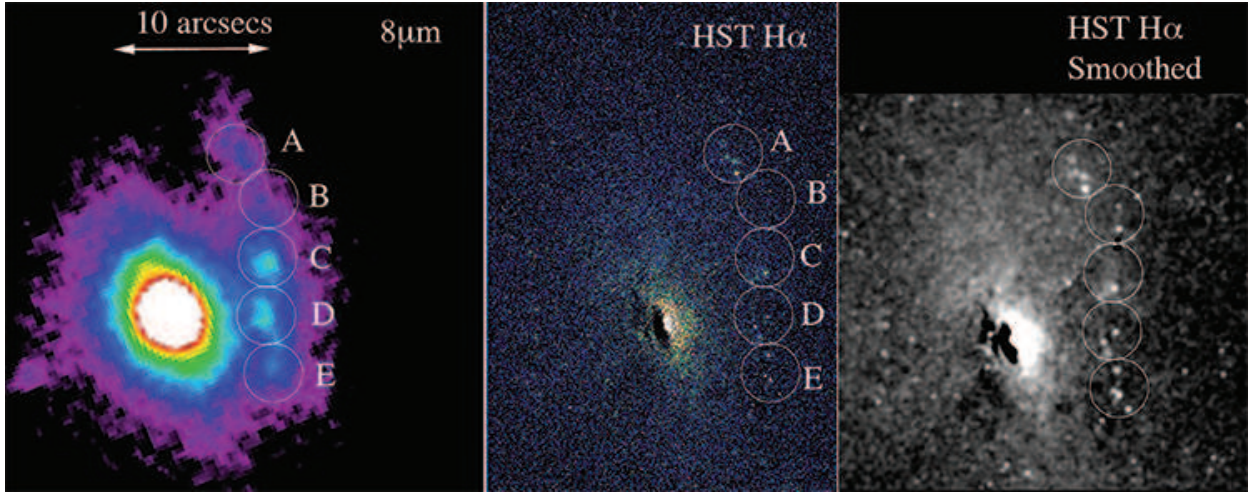}
\caption{Left: The 8$\mu$m IRAC image of NGC 3226 showing areas selected for photometric analysis, A thru E (See text). Middle: H$\alpha$ image of NGC 3226 taken with HST and showing a series of tiny HII regions associated with the IRAC blobs. Right: The same H$\alpha$ image smoothed with a Gaussian to bring out the fainter extended emission. 
In addition to the H$\alpha$ regions, significant extended emission is seen, as well as strong dust-absorption to the south of the nucleus. 
}
\end{figure}

\begin{figure}[!ht]
\vspace{0.3 in}
\plotone{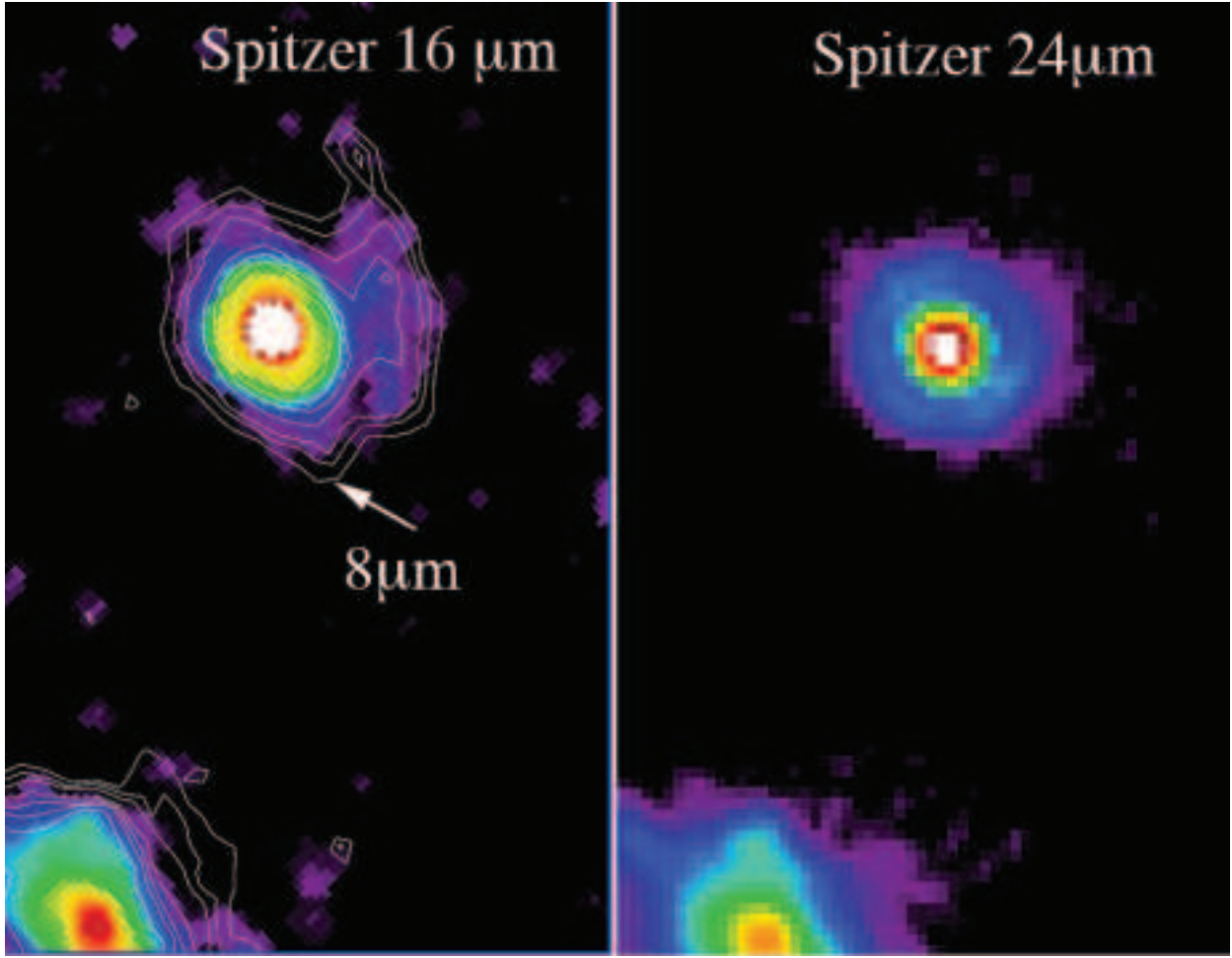}
\caption{Left: Contours of 8$\mu$m IRAC emission superimposed on a
16$\mu$m IRS Peak-up image of the galaxy. Note the close association
between the 8$\mu$m contours and the plume. Right: The 24$\mu$m image of the same region. Note the absence of dust emission from the plume. This suggests
that the 16$\mu$m emission from the plume contains either strong 0-0S(1) H$_2$ emission at 17$\mu$m, or 16$\mu$m PAH emission, or both. IRS spectra, 
which do not cover this region, show unusually strong H$_2$ emission from the center of the galaxy.  
}
\end{figure}

\begin{figure}[!ht]
\vspace{0.3in}
\plotone{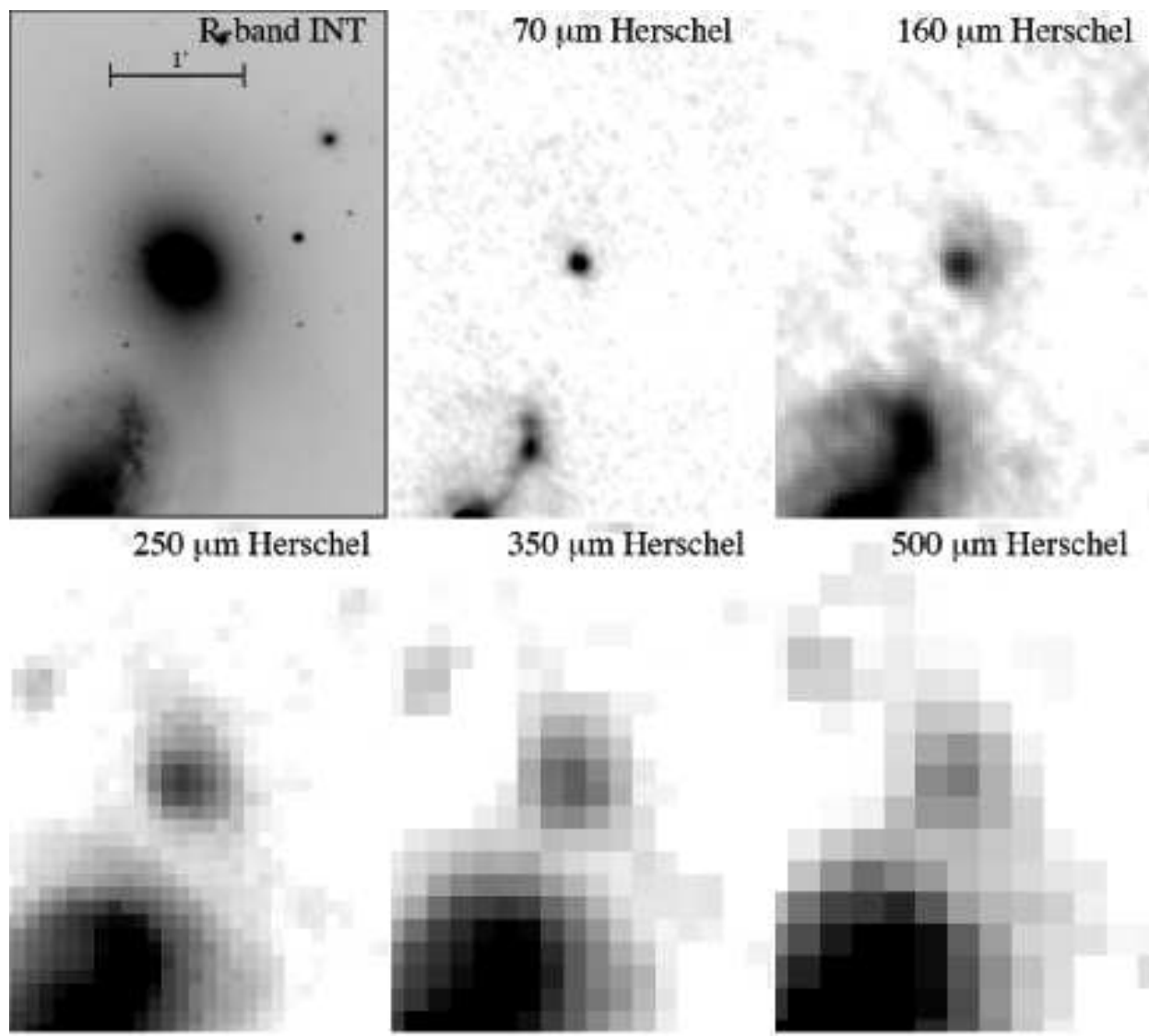}
\caption{A mosaic of images showing emission from NGC 3226 as seen by
the PACS and SPIRE photometers onboard {\it Herschel}. The left-hand top panel
shows the galaxy on the same scale in R-band. Note the very compact nature
of the emission at 70$\mu$m,and possible extended emission to the north-west
of NGC3226 at 160$\mu$m and at longer wavelengths, perhaps due to dust associated with the base of the HI filament shown in previous figures.
The spatial resolution of the five {\it Herschel} photometric maps are
x,y,z,a,b respectively. The maps were made using {\it Scanamorphos} \citep[see][]{rou13}.            
}
\end{figure}

\begin{figure}[!ht]
\vspace{0.3in}
\plotone{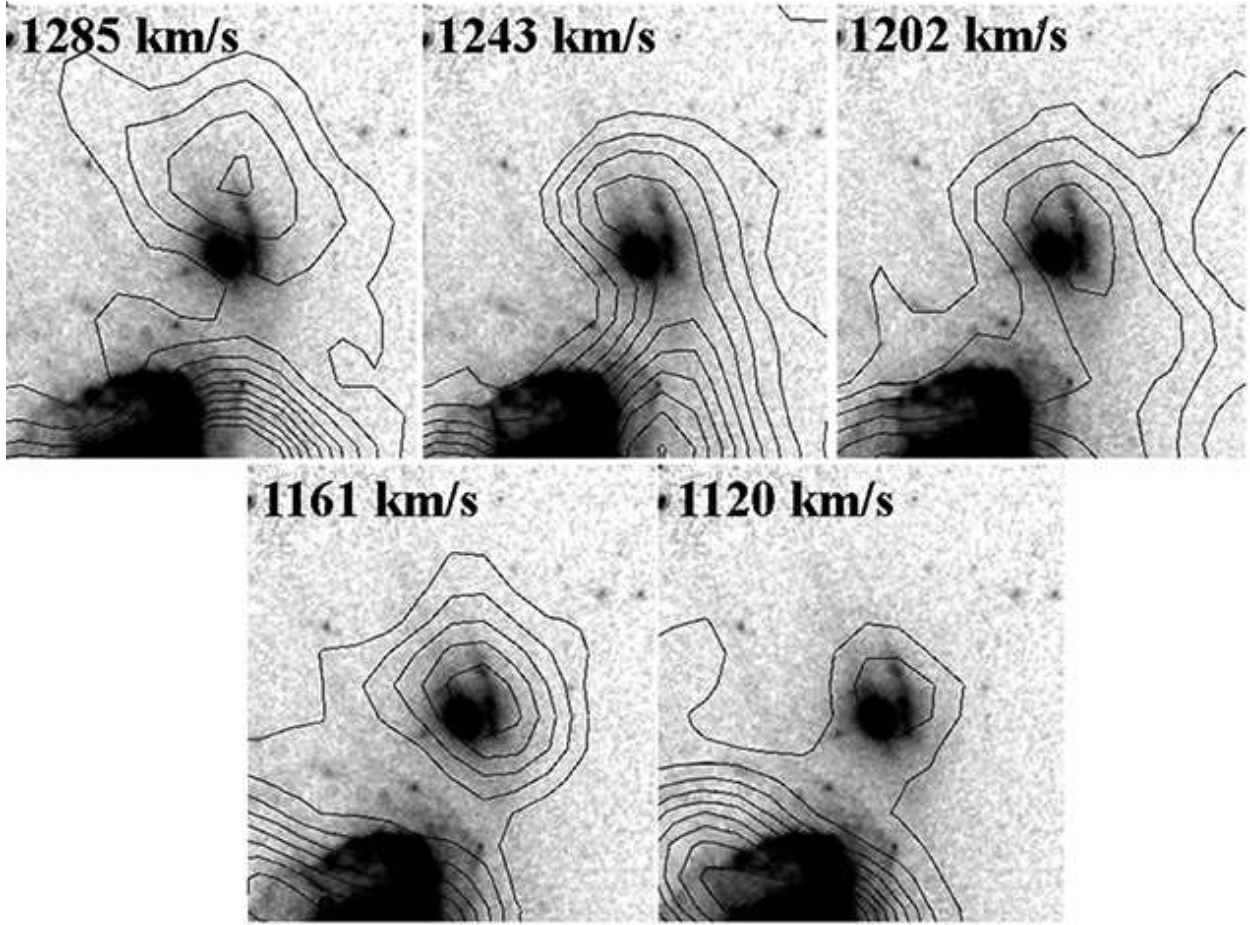}
\caption{Neutral hydrogen emission from 40km s$^{-1}$ wide channels
(contours) superimposed on the IRAC 8$\mu$m emission (greyscale) of
NGC 3226. The figure shows data obtained from Mundell (1995) D-array
VLA data. The sequence starts at 1285 km s$^{-1}$ (heliocentric velocity),
and shows that as the velocity decreases there is a clear association
of the HI centroid with the 8$\mu$m emitting PAH emission. Notice how,
at the lowest velocity, the HI terminates in the 8$\mu$m plume and
does not extend beyond the base of that plume--suggesting that the
northern HI plume actually ends at the 8$\mu$m feature. Channel maps
showing HI at even higher velocities than shown here continue up into
the much more extended northern HI plume of NGC 3226 (Mundell
1995).}
\end{figure}

\begin{figure}[!ht]
\vspace{0.3in}
\plotone{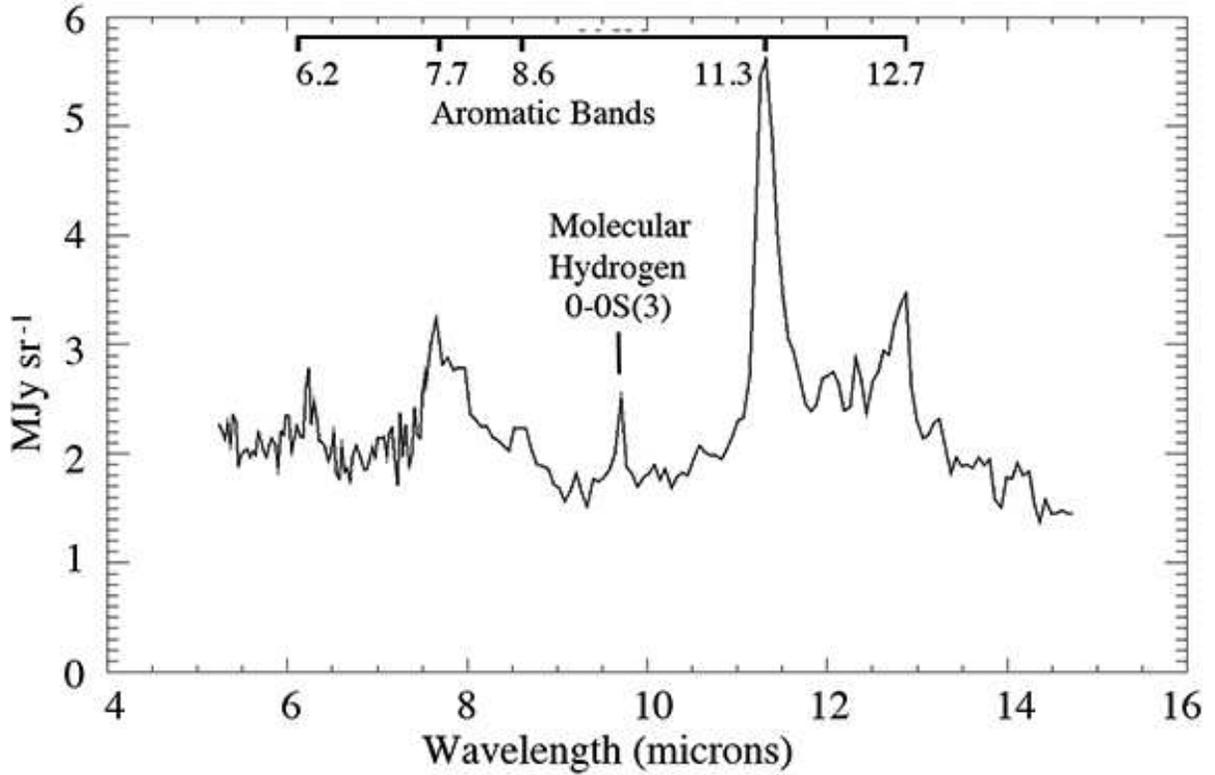}
\caption{Spectrum of NGC 3226  made with the short-low (SL) module of the {\it Spitzer IRS} extracted from a 22 x 16 arcsec$^2$ area in the center of a small map of NGC 3226. In addition to the broad PAH features at 6.2, 7.7, 8.6 11.3 and 12.7 $\mu$m, molecular hydrogen is detected in the 0-0S(3)9.66$\mu$m line.}
\end{figure}

\begin{figure}[!ht]
\vspace{0.3in}
\plotone{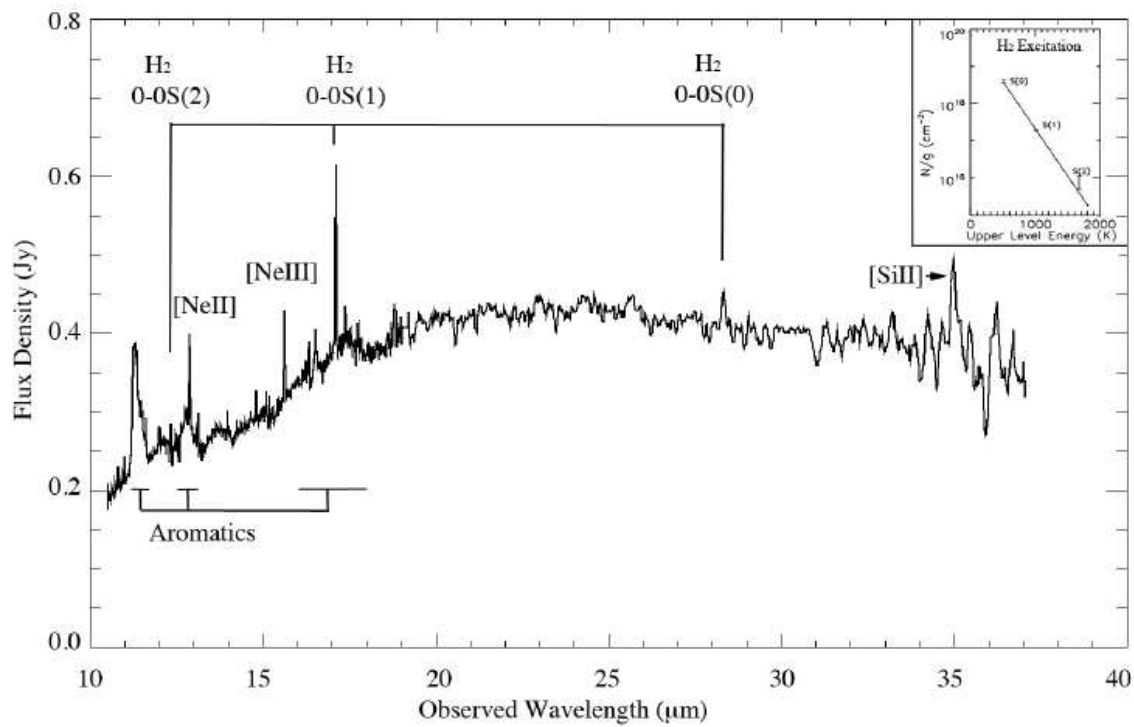}
\caption{High resolution (SH and LH) IRS spectrum of the center of NGC3226 showing clear
evidence of strong molecular hydrogen, PAH emission, and other emission lines.
 Note the flattening of the continuum beyond 20$\mu$m, typical
of low-luminosity AGN (Ogle et al. (2010)). The inset shows the excitation diagram for molecular hydrogen, and a single temperature fit
to the points, under the assumption of fully extended H$_2$ emission on the scale of the LH IRS slit indicating a temperature for the H$_2$ of T = 144 K. However, a more realistic assumption is that it is of intermediate scale (10-15 arcsecs) and the temperature under these assumptions is found to be lower (T=124 K, see text).  Note that there is an unknown
contribution to the continuum by zodiacal light which was not removed from the spectrum. 
}
\end{figure}

\begin{figure}[!ht]
\vspace{0.3in}
\plotone{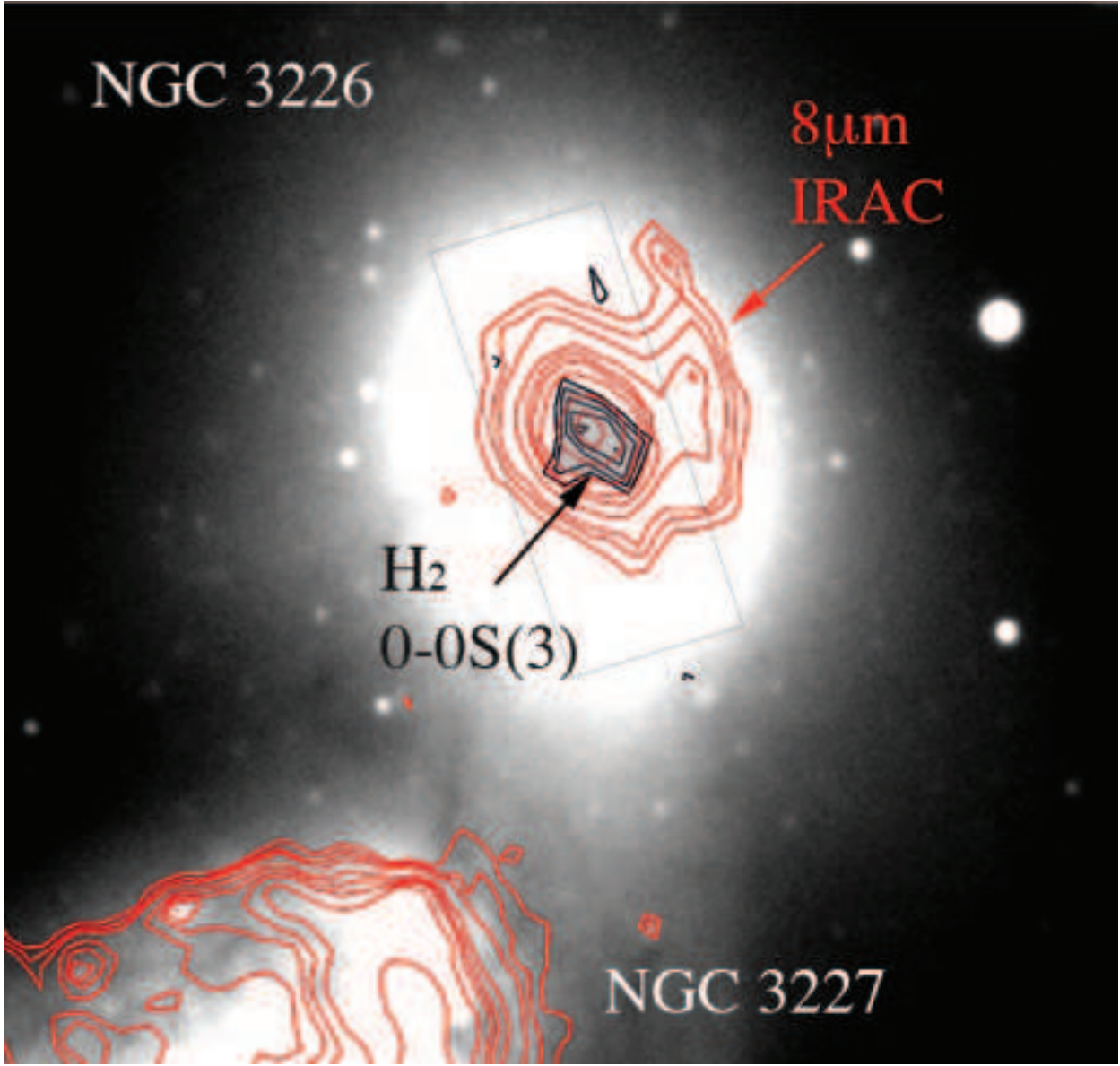}
\caption{Contours of 8$\mu$m IRAC emission (red), and 0-0S(3)9.66$\mu$m
molecular hydrogen (black/grey-filled) superimposed on a B-band
visible-light CCD image of NGC3226 (and NGC 3227 to the South). The
grey box shows the area covered in the spectra mapping of NGC3226 by
the short-low {\it Spitzer} IRS instrument. Note that the area mapped
did not include the plume feature.
Details of the mapping are given in
the main text.
}
\end{figure}

\newpage

\begin{figure}[!ht]
\centering
\includegraphics[width=1.0\textwidth]{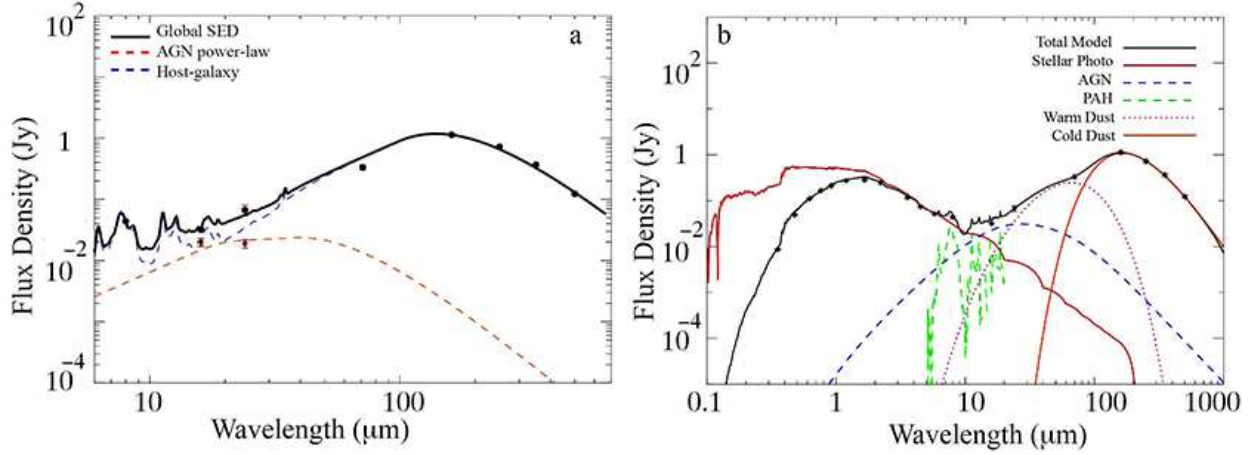}
\caption{ a) A fit (solid black line) to the mid and Far-IR SED of NGC 3226 using the method of Mullaney et al. (2011) which uses empirically defined mid- and far-IR spectral templates to model the host galaxy (blue dotted line), and a broken power-law model for the AGN (red dotted line). We find the best fit AGN contributes 20$\pm$5$\%$ of the host galaxy's IR luminosity. Black points show the Spitzer and Herschel data for the global galaxy (Table 4).  Red data points (not used in the fit, but extracted from a nuclear aperture) show good agreement with the broken power-law AGN component determined from the global fit, b)  SED modeling using the Sajina et al. (2012) approach (see text) which models the full UV-far-IR SED.  This approach includes models for a young stellar population (red solid line), a power-law AGN (blue dashed line), as well as two dust components (red dotted and orange solid line) + a PAH contribution (green line), and assumes a fraction of the UV is re-radiated in the IR. The black solid line shows the model in comparison with these data points from Table 4.  The AGN luminosity compared with the IR dust luminosity is similar to the previous modeling (18$\pm$4$\%$), but additionally allows several key parameters of the host galaxy to be quantified (see Table 5). }
\end{figure}

\begin{figure}[!ht]
\vspace{0.3in}
\plotone{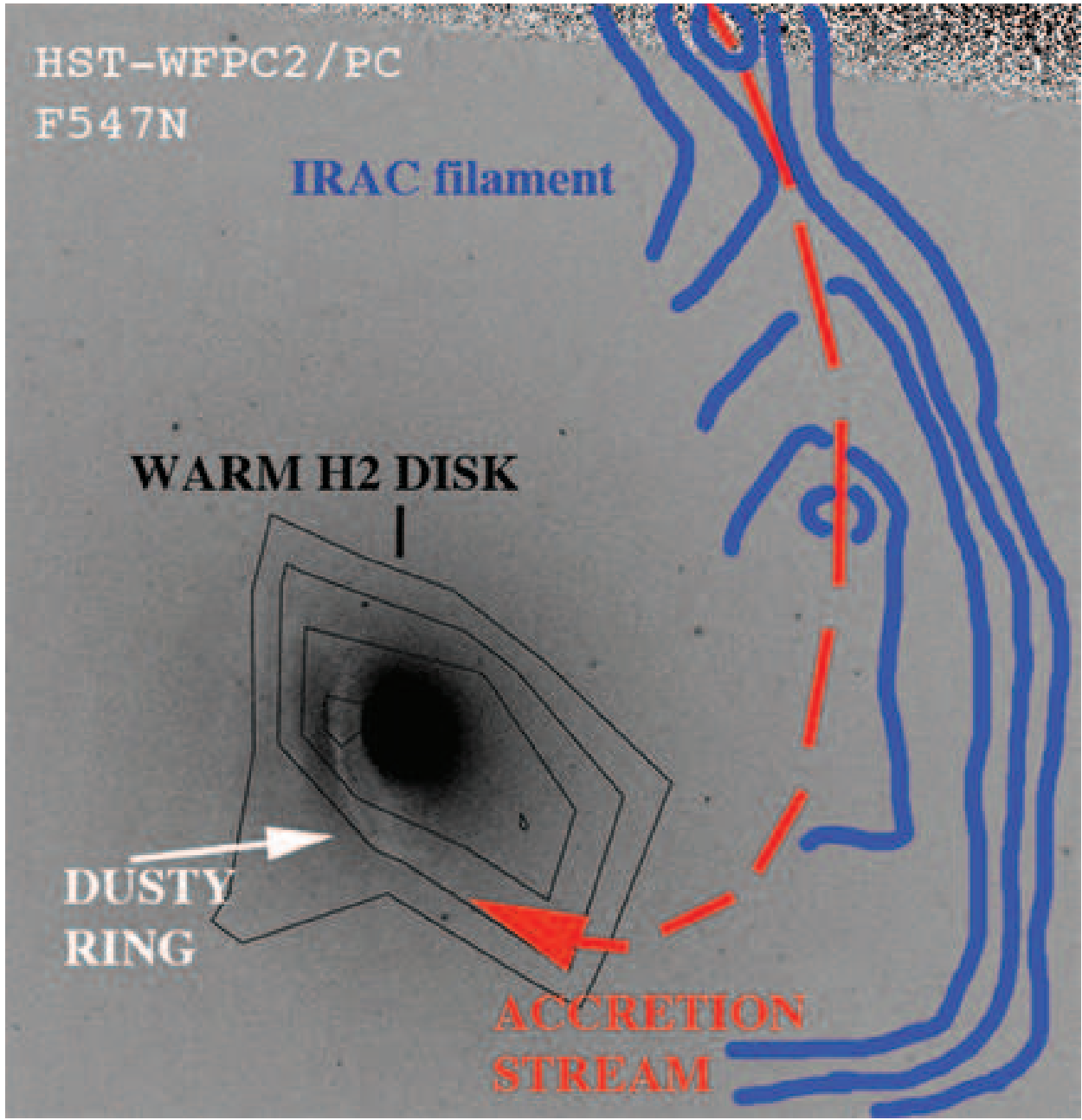}
\caption{Greyscale image of the core of NGC 3226 showing a partial ring of dust approximately coincident with the warm H$_2$ emission (black contours). The blue 
contours show the position of the 8$\mu$m filament which lies at the base of the HI plume. The red dotted line is a representation of the possible
flow of material from the filament onto the galaxy.}
\end{figure}

\begin{figure}[!ht]
\includegraphics[width=3in]{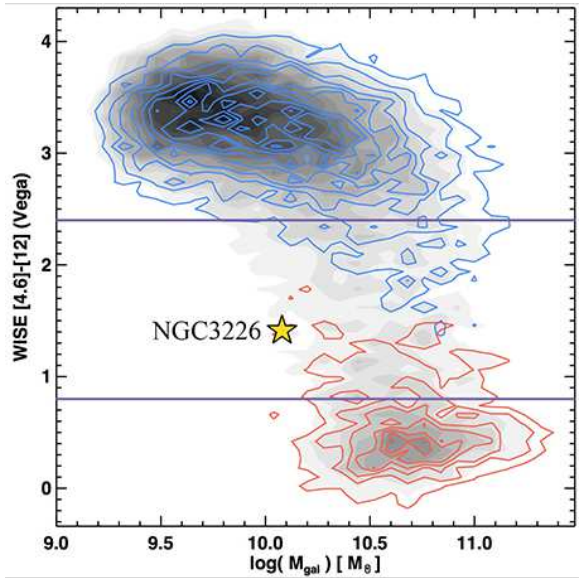}
\caption{WISE color-mass diagram from Alatalo et al. (2014) based on a sample of spectroscopically-selected SDSS galaxies (colored contours), and  galaxies from  the "Galaxy Zoo" of Schawinski et al. (2014; underlying grey scale). The figure shows the clear bifurcation of early-type (red contours) and late-type (blue contours) galaxies. Contours are 5$\%$ steps of the total galaxy population studied. The area between the two purple horizontal lines in the diagram shows the InfraRed Transition Zone (IRTZ). The galaxies in the IRTZ are shown spectroscopically to be dominated by strong LINERs, a population of Seyferts,  and shocked post-starburst galaxy candidates (Alatalo et al. 2014b).  NGC 3226 clearly falls in the IRTZ,  emphasizing its similarity to a larger body of potentially evolving galaxies (see text).}
\end{figure}

\end{document}

\tablenotetext{ }{FUV, NUV from GALEX and   16$\mu$m, 30$\mu$m from Spitzer IRS} compared with Lanz et al. (2013)} 

A recent study using the {\it Spitzer} InfraRed Spectrograph (IRS) showed that more than 10$\%$
of Hickson Compact Group galaxies show unusually large ratios of warm
molecular hydrogen emission to PAH emission (poly-cyclic aromatic hydrocarbon)--symptomatic of galaxies with strongly shocked gaseous
media \citep{clu13}. Rather than showing a wide range of colors, these
galaxies fall mainly in region of mid-IR color space which Cluver et al. called 
the "mid-IR green-valley", as has previously been associated with a dearth of galaxies in this region \citep{joh07, wal10}.  This region contains a significant number of
UV-optically defined green-valley galaxies after correction for internal extinction \citep{bit10,bit11}.  
Could these galaxies
lie in the green-valley because something is causing the gas to be
highly turbulent, inhibiting star formation and causing gas
rich-systems to drift from the blue cloud? One characteristic of these
galaxies is that they tend to lie in compact groups containing large
numbers of early-type galaxies. NGC 3226, the subject of the current paper, shows some similarities
with the H$_2$-excess HCG galaxies, perhaps because it is part of a multi-galaxy Arp 94 system.